\documentclass[journal]{IEEEtran}
\IEEEoverridecommandlockouts
\usepackage{cite}
\usepackage{amsmath,amssymb,amsfonts}
\usepackage{graphicx}
\usepackage{textcomp}
\usepackage{xcolor}
\usepackage{algorithm}
\usepackage{algorithmic}
\usepackage{enumitem}
\usepackage{footnote}
\usepackage{multirow}
\usepackage{threeparttable}
\usepackage{makecell}
\usepackage{subfigure}
\usepackage{bm}
\ifCLASSOPTIONcompsoc
\usepackage[caption=false, font=normalsize, labelfont=sf, textfont=sf]{subfig}
\else
\usepackage[caption=false, font=footnotesize]{subfig}
\fi

\usepackage{array}

\def\BibTeX{{\rm B\kern-.05em{\sc i\kern-.025em b}\kern-.08em
    T\kern-.1667em\lower.7ex\hbox{E}\kern-.125emX}}

\usepackage{etoolbox}
\let\mybibitem\bibitem
\renewcommand{\bibitem}[1]{%
  \ifstrequal{#1}{nature}
    {\color{blue}\mybibitem{#1}}
    {\color{black}\mybibitem{#1}}%
}

\begin{document}

\title{A Low-Overhead Incorporation-Extrapolation based Few-Shot CSI Feedback Framework for Massive MIMO Systems}

\author{Binggui Zhou,
        Xi Yang,
        Jintao Wang,
        Shaodan Ma,
        Feifei Gao,
        and Guanghua Yang
\thanks{This work was supported in part by the National Natural Science Foundation of China under Grants 62261160650, 62171201, and 62301221; in part by the Major Talent Program of Guangdong Provincial under Grant 2019QN01S103; in part by the Shanghai Pujiang Program under Grant 22PJ1403100; in part by the open research fund of National Mobile Communications Research Laboratory, Southeast University (No. 2024D02); in part by the Science and Technology Development Fund, Macau SAR, under Grants 0087/2022/AFJ and 001/2024/SKL; and in part by the Research Committee of University of Macau under Grants MYRG-GRG2023-00116-FST-UMDF and MYRG2020-00095-FST. (Corresponding authors: Xi Yang; Guanghua Yang.)}
\thanks{Binggui Zhou is with the School of Intelligent Systems Science and Engineering, Jinan University, Zhuhai 519070, China; and also with the State Key Laboratory of Internet of Things for Smart City and the Department of Electrical and Computer Engineering, University of Macau, Macao 999078, China (e-mail: binggui.zhou@connect.um.edu.mo).}
\thanks{Xi Yang is with the School of Communication and Electronic Engineering, East China Normal University, Shanghai 200241, China, and is also with the National Mobile Communications Research Laboratory, Southeast University, Nanjing 210096, China (email: xyang@cee.ecnu.edu.cn).}
\thanks{Jintao Wang and Shaodan Ma are with the State Key Laboratory of Internet of Things for Smart City and the Department of Electrical and Computer Engineering, University of Macau, Macao 999078, China (e-mails: wang.jintao@connect.um.edu.mo, shaodanma@um.edu.mo).}
\thanks{Feifei Gao is with the Institute for Artificial Intelligence, Tsinghua University (THUAI), State Key Lab of Intelligent Technologies and Systems, Tsinghua University, Beijing National Research Center for Information Science and Technology (BNRist), and Department of Automation, Tsinghua University, Beijing 100084, China (e-mail: feifeigao@ieee.org).}
\thanks{Guanghua Yang is with the School of Intelligent Systems Science and Engineering and the Guangdong International Cooperation Base of Science and Technology for GBA Smart Logistics, Jinan University, Zhuhai 519070, China (e-mail: ghyang@jnu.edu.cn).}
}

\maketitle

\begin{abstract}
Accurate channel state information (CSI) is essential for downlink precoding in frequency division duplexing (FDD) massive multiple-input multiple-output (MIMO) systems with orthogonal frequency-division multiplexing (OFDM). However, obtaining CSI through feedback from the user equipment (UE) becomes challenging with the increasing scale of antennas and subcarriers and leads to extremely high CSI feedback overhead. Deep learning-based methods have emerged for compressing CSI but these methods generally require substantial collected samples and thus pose practical challenges. Moreover, existing deep learning methods also suffer from dramatically growing feedback overhead owing to their focus on full-dimensional CSI feedback. To address these issues, we propose a low-overhead  \underline{I}ncorporation-\underline{E}xtrapolation based \underline{F}ew-\underline{S}hot CSI feedback \underline{F}ramework (IEFSF) for massive MIMO systems. An incorporation-extrapolation scheme for eigenvector-based CSI feedback is proposed to reduce the feedback overhead. Then, to alleviate the necessity of extensive collected samples and enable few-shot CSI feedback, we further propose a knowledge-driven data augmentation (KDDA) method and an artificial intelligence-generated content (AIGC) -based data augmentation method by exploiting the domain knowledge of wireless channels and by exploiting a novel generative model, respectively. Experimental results based on the DeepMIMO dataset demonstrate that the proposed IEFSF significantly reduces CSI feedback overhead by 64 times compared with existing methods while maintaining higher feedback accuracy using only several hundred collected samples.
\end{abstract}

\begin{IEEEkeywords}
Massive MIMO, CSI Feedback, Few-Shot Learning, Domain Knowledge, AIGC
\end{IEEEkeywords}

\section{Introduction}

\IEEEPARstart MASSIVE multi-input multi-output (MIMO) technology is an essential technology for the fifth generation (5G) wireless communication systems and beyond \cite{larsson2014massive, feng2018impact, feng2019twoway, tan2022antenna, yang2023antenna, zhou2024pay}. To fulfill the potential of massive MIMO, accurate channel state information (CSI) should be acquired no matter at the time division duplexing (TDD) systems or at the frequency division duplexing (FDD) systems. TDD systems allow the base station (BS), which is equipped with a large number of antennas, to obtain downlink CSI by inferring from the estimated uplink CSI via the reciprocity between uplink and downlink channels. While in FDD systems, the BS needs to attain downlink CSI through CSI feedback from the user equipment (UE) since the uplink-downlink channel reciprocity only partially holds.

Although there are conventional vector quantization-based and codebook-based CSI feedback methods for FDD MIMO systems, both of them face significant challenges in FDD massive MIMO systems\cite{love2008overview}, especially under the wideband bandwidth configuration. As the number of antennas at the BS in massive MIMO systems increasingly increases and the operational bandwidth progressively widens, the dimension of the CSI matrix grows significantly, leading to unaffordable high feedback overhead. To reduce the feedback overhead, in recent years, numerous deep learning-based CSI feedback methods have been proposed to compress and feedback the CSI matrix at the UE and then reconstruct it at the BS, including CsiNet \cite{wen2018deep}, CsiNet+\cite{guo2020convolutional}, SRNet\cite{chen2022highaccuracy}, TransNet \cite{cui2022transnet}, etc. CsiNet, known as the first deep learning-based CSI feedback method, was proposed to compress and recover CSI by learning the channel structure from training samples with convolutional neural networks (CNNs). In \cite{guo2020convolutional}, CsiNet+ was developed from CsiNet by considering multiple-rate bit-level compressive sensing for CSI feedback. SRNet in \cite{chen2022highaccuracy} employed the image super-resolution technology to exploit CSI frequency characteristics. In \cite{cui2022transnet}, TransNet, a neural network for CSI feedback based on the Transformer architecture \cite{vaswani2017attention}, was presented. TransNet can learn connections between angular domain elements of the CSI and thus showed significant performance improvement compared to the existing deep learning-based CSI feedback methods. Note that, to efficiently reduce the CSI feedback overhead, all these aforementioned deep learning-based CSI feedback methods are typically designed with an encoder-decoder architecture, i.e., the UE-side encoder compresses the CSI matrix and then feedbacks the compressed CSI to BS through the feedback link, and the BS-side decoder reconstructs the CSI matrix after receiving the compressed one. In addition, the CSI feedback efficiency can be further promoted by exploiting the channel property. For example, in \cite{wang2021compressive}, the SampleDL method was proposed via capitalizing on channel correlations in both temporal and frequency domains. Based on the channel correlations, only a portion of the CSI elements was selected and a much size-reduced massive MIMO CSI matrix was obtained before the CSI feedback. This, therefore, effectively reduced the feedback overhead and computational complexity on the UE side. Nonetheless, such downsampling process will inevitably lead to information loss due to the direct dropping of the orthogonal frequency-division multiplexing (OFDM) symbols and the subcarriers.

Furthermore, previous research on deep learning-based CSI feedback has primarily focused on full channel feedback, which involves compressing the full downlink channel matrix on the UE side and subsequently reconstructing it on the BS side after feedback. In contrast, as specified in the current 3rd Generation Partnership Project (3GPP) specification\cite{3gpp2020ts38214}, the commonly used codebook-based CSI feedback is based on an implicit CSI feedback scheme. That is, only partial channel information (e.g., the precoding matrix indicator (PMI)) is encoded with an encoding module at the UE, fed back to the BS, and decoded with a decoding module at the BS. In view of the increasing antenna scale and transmission bandwidth, CSI feedback based on PMI will be greatly helpful in reducing the feedback overhead. As a result, eigenvector-based CSI feedback methods were proposed in \cite{chen2022deep,liu2021evcsinet,shen2022mixernet}, where eigenvectors were used as the PMI for downlink precoding.

{Despite the effectiveness of these deep learning-based CSI feedback methods, a large number of collected samples (e.g., tens of thousands of samples) are generally required during the training period to obtain deep learning models with strong generalization ability. For example, in \cite{wen2018deep}, $100,000$ collected samples were required to train the CsiNet under the system configuration of $32$ antennas and $1024$ subcarriers. In practice, collecting such large amounts of samples necessitates substantial CSI acquisition time and pilot training overhead \cite{guo2024deep}, which is always unaffordable. In addition, confined by the generalization capability of existing DL-based CSI feedback methods, it is also necessary for these methods to collect new data for new scenarios, making the extensive data collection impractical for employment. Moreover, UE mobility poses another challenge to data collection, i.e., the fast-changing channel conditions originating from the mobility necessitate much more time-frequency resources to obtain accurate CSI samples. To circumvent such data collection problems, learning from a merely limited number of collected samples is highly advocated, which is known as few-shot learning (FSL) in machine learning. To realize FSL, effectively augmenting the limited collected samples is an essential approach.} {Existing data augmentation methods that have been used for CSI feedback can be roughly categorized into four categories, i.e., the image processing-based, the ray-tracing based, the knowledge-driven, the artificial intelligence-generated content (AIGC) -based data augmentation methods.} The image processing-based methods regard each CSI matrix from the CSI samples as an image with two channels (typically the real and the imaginary parts of the complex-valued CSI matrix). Therefore, traditional image data augmentation methods, e.g., scaling, shifting, and rotation, can be applied to CSI samples directly\cite{xiao2022ai}. However, without proper guidance and evaluation, such image processing-based data augmentation methods may destroy the original structural information of CSI samples (e.g., the orientation of the antenna array), which may introduce nuisances contaminating the model training. {Recently, ray-tracing based data augmentation methods have shown the potential to augment CSI samples by utilizing digital twin techniques\cite{jiang2024digital,alkhateeb2023realtime}. However, ray-tracing based data augmentation methods essentially require a digital reconstruction of exact wireless communication environments to make the generated CSI samples similar to the measured CSI samples (i.e., following the same distribution), which is significantly challenging and computationally intensive in practice. Inaccurate digital reconstruction might produce dissimilar samples and ultimately degrade the model performance. In particular, for mobile scenarios, it is even impossible to accurately model the mobility of mobile users, making ray-tracing based data augmentation methods less beneficial.} The knowledge-driven data augmentation methods aim to explore domain knowledge of wireless channels to generate augmented samples. For example, in \cite{liu2022training}, two knowledge-driven data augmentation methods were proposed by considering the geographical continuity of antennas and the delay property of MIMO channels. Recently, with the emergence of ChatGPT\cite{openai2022introducing}, Stable Diffusion\cite{ramesh2022hierarchical}, DALL-E-2\cite{rombach2022highresolution}, etc., AIGC, which indicates content generated with generative models, has been garnering increasing attention from both academia and industry \cite{cao2023comprehensive, wu2023aigenerated}. Exploiting AIGC for data augmentation becomes promising in various fields\cite{galashov2022data, chen2023unified, yang2023aigenerated}. Specifically, generative adversarial networks (GANs), one kind of generative model for AIGC, have been studied for data augmentation in wireless communications. GANs first model wireless channels by learning the distribution from a set of channel realizations\cite{yang2019generativeadversarialnetworkbased, ye2020deep,xiao2022channelgan}. After channel modeling, the trained GANs could generate various channel realizations following the same distribution to enhance the performance and generalization ability of deep learning-based CSI feedback methods\cite{xiao2022channelgan}. Nevertheless, training a GAN typically requires large amounts of training data\cite{li2020fewshot}, which contradicts the intention of exploiting GAN for data augmentation, especially under few-shot scenarios. Therefore, a practical AIGC-based data augmentation method for CSI feedback deserves further exploration.

Hence, in this paper, we consider low-overhead eigenvector-based CSI feedback for massive MIMO systems under few-shot scenarios. To simultaneously reduce feedback overhead and computational complexity, we first propose the incorporation-extrapolation CSI feedback scheme by exploiting frequency domain correlations. Then, to alleviate the necessity of the extensive collected samples and enable few-shot CSI feedback, we further propose a knowledge-driven data augmentation method and an AIGC-based data augmentation method. We refer to the entire framework as the \underline{I}ncorporation-\underline{E}xtrapolation based \underline{F}ew-\underline{S}hot CSI feedback \underline{F}ramework (\textbf{IEFSF}), which combines the incorporation-extrapolation CSI feedback scheme with domain knowledge and AIGC data augmentation to enable low-overhead few-shot massive MIMO CSI feedback.

The major contributions of this paper are summarized as follows:

\begin{enumerate}

    \item We propose the incorporation-extrapolation CSI feedback scheme to simultaneously reduce feedback overhead and computational complexity. Unlike \cite{wang2021compressive}, which directly downsamples the frequency domain subcarriers, the proposed incorporation-extrapolation scheme first yields a low-dimensional eigenvector-based CSI matrix through the incorporation process by dividing all subcarriers into several groups and generating eigenvectors for all groups. Then, a Transformer-based CSI compression and reconstruction network is proposed for CSI compression and reconstruction by exploiting the channel correlation among multiple subcarrier groups. Next, a novel lightweight frequency extrapolation network is proposed to recover the full-dimensional eigenvector-based CSI matrix on the BS side with low computational complexity by exploiting multi-path extrapolation and residual learning.\footnote{Note that the `multi-path' extrapolation indicates extrapolation through multiple extrapolation modules, which differs from the concept of `multipath' propagation phenomenon in wireless signal propagation.} {The proposed extrapolation network significantly reduces the computational complexity of frequency domain extrapolation compared with existing extrapolation networks.}
    
    \item We further propose a knowledge-driven data augmentation method and an AIGC-based data augmentation method to enable few-shot CSI feedback. The proposed knowledge-driven data augmentation method exploits frequency domain correlations at various granularities, which is easy to implement and computationally efficient. To further promote the effectiveness of data augmentation, we propose the Extrapolation Generative Adversarial Network (EGAN), a novel generative model for AIGC, to generate diverse training samples. Unlike existing generative models that focus on modeling the channels, our EGAN learns to directly generate eigenvector-based CSI matrices, offering significant computational efficiency advantages. The knowledge-driven data augmentation method is applicable for scenarios with very few collected samples (e.g., only $100$ - $200$ samples) while the AIGC-based method exhibits better performance when there are several hundred samples.

    \item We evaluate the performance of the proposed IEFSF under both indoor and outdoor scenarios. Numerical results indicate that both the proposed knowledge-driven data augmentation method and the proposed AIGC-based data augmentation method exhibit excellent performance in achieving accurate CSI feedback with only several hundreds of collected samples under the system configuration of $32$ antennas and $100$ MHz bandwidth with $1024$ subcarriers. In addition, the proposed IEFSF can achieve a higher CSI feedback accuracy than existing CSI feedback methods but with $64$ times less feedback overhead.
    
\end{enumerate}

The remainder of this paper is organized as follows. In Section \ref{Sec. Sys.}, we introduce the system model and the CSI feedback procedure. In Section \ref{Sec. IE}, we propose the incorporation-extrapolation CSI feedback scheme. In Section \ref{Sec. DataAug}, we propose the knowledge-driven data augmentation method and the AIGC-based data augmentation method to enable few-shot CSI feedback. In Section \ref{Sec. Exp.}, extensive simulation results and computational complexity analysis are presented. Finally, we conclude this work in Section \ref{Sec. Con.}.

\textit{Notation}: Underlined bold uppercase letters, bold uppercase letters, and bold lowercase letters represent tensors, matrices, and vectors, respectively. Calligraphy uppercase letters represent sets. $\odot$, $\mathbb{E}\{\cdot\}$, $\operatorname{Var}\{\cdot\}$, and $\|\cdot\|_2$ denote the Hadamard product, expectation, variance, and L2 norm, respectively. $\inf (\cdot)$ and $\sup (\cdot)$ denote the infimum and supremum, respectively. $\gamma(x, y)$ denotes the joint distribution of $x$ and $y$. $C_a^b$ denotes the combination operator indicating choosing $b$ elements from $a$ elements.

\section{System Model and CSI Feedback Procedure}\label{Sec. Sys.}

\subsection{System Model}

In this paper, we consider a single-cell FDD massive MIMO system, where a single BS equipped with $N_T\gg1$ antennas serves a single UE equipped with $N_R$ antennas.\footnote{Note that a single-user massive MIMO system is considered in this paper, but the proposed framework can be directly applied to multi-user scenarios.} The system operates with the orthogonal frequency division multiplexing (OFDM) modulation with a total of $N_c$ subcarriers. In the downlink transmission, the received signal at the $i$-th subcarrier can be represented as
\begin{equation}\label{signal}
    \mathbf{y}_i = \mathbf{H}_i \mathbf{P}_i \mathbf{s}_i + \mathbf{n}_i,
\end{equation}
where $\mathbf{H}_i \in \mathbb{C}^{N_R \times N_T}$, $\mathbf{P}_i \in \mathbb{C}^{N_T \times N_s}$, $\mathbf{s}_i \in \mathbb{C}^{N_s \times 1}$, and $\mathbf{n}_i \in \mathbb{C}^{N_R \times 1}$ correspond to the downlink channel, the downlink precoding matrix, the signal to be sent, and the noise, at the $i$-th subcarrier, respectively. And $\mathbf{P}_i = [\mathbf{p}_{i,1}, \mathbf{p}_{i,2}, \ldots, \mathbf{p}_{i,N_{s}}]$, where $\mathbf{p}_{i,j} \in \mathbb{C}^{N_T \times 1}$, $j = 1,2, \ldots, N_s$, are the column vectors of $\mathbf{P}_i$ and $N_s$ is the number of data streams with $N_s \leq N_R$.

To maximize system performance, the downlink precoding matrix $\mathbf{P}_i$ needs to be designed based on the downlink channel $\mathbf{H}_i$. However, in FDD systems, the BS cannot directly derive $\mathbf{H}_i$ from the uplink channel information obtained through uplink pilot training. In addition, the increasing number of antennas and subcarriers leads to a significant growth of the channel dimension, making the overhead of full channel feedback become prohibitive. As a result, an implicit feedback mechanism (e.g., PMI feedback) is advocated by current 3GPP specifications. In this paper, we consider eigenvalue decomposition (EVD) -based downlink precoding and thus the eigenvector-based CSI feedback is adopted as in \cite{chen2022deep,liu2021evcsinet,shen2022mixernet}. Specifically, denote the Gram matrix of $\mathbf{H}_i$ as $\mathbf{R}_i \in \mathbb{C}^{N_T \times N_T}$, i.e.,
\begin{equation}\label{correlation_mat}
    \mathbf{R}_i = \mathbf{H}_i^{H}\mathbf{H}_i,
\end{equation}
where $\mathbf{H}_i^{H}$ is the Hermitian of $\mathbf{H}_i$. The corresponding dominant eigenvector $\mathbf{w}_i \in \mathbb{C}^{N_T \times 1}$ of $\mathbf{R}_i$ can be calculated by EVD as\footnote{To simplify the following descriptions, we only consider the feedback of the dominant eigenvector of $\mathbf{R}_i$ in this paper, but multiple eigenvectors can also be fed back with the proposed framework.}
\begin{equation}\label{eigen}
\mathbf{R}_i \mathbf{w}_i = \lambda^\text{max}_i \mathbf{w}_i,
\end{equation}
where $\lambda^\text{max}_i$ is the maximum eigenvalue of $\mathbf{R}_i$.

By concatenating the dominant eigenvectors corresponding to all $N_{c}$ subcarriers, we have the eigenvector-based CSI matrix $\mathbf{W} \in \mathbb{C}^{N_T \times N_{c}}$ as
\begin{equation}\label{whole_eigen}
\mathbf{W} \triangleq [\mathbf{w}_1, \mathbf{w}_2, \ldots, \mathbf{w}_{N_{c}}].
\end{equation}

Then, the eigenvector-based CSI matrix $\mathbf{W}$, instead of the whole downlink channel $\mathbf{H} \triangleq [\mathbf{H}_1, \mathbf{H}_2, \ldots, \mathbf{H}_{N_{c}}] \in \mathbb{C}^{N_R \times (N_T \times N_{c})}$, will be fed back to the BS, and the $i$-th column vector of the reconstructed eigenvector-based CSI matrix $\hat{\mathbf{W}}$ will serve as the downlink precoding vector $\mathbf{p}_i$ at the $i$-th subcarrier.

\subsection{CSI Feedback Procedure}
In most existing deep learning-based CSI feedback methods, after obtaining $\mathbf{W}$, the UE starts the CSI feedback process by first compressing $\mathbf{W}$ into the CSI feature vector $\mathbf{v} \in \mathbb{R}^{L_q \times 1}$ via the compression module, then quantizing $\mathbf{v}$ to the bit stream via the quantization module, and finally sending the bit stream to the BS via the feedback link. Once receiving the bit stream, the BS first dequantizes the bit stream to the CSI feature vector $\hat{\mathbf{v}} \in \mathbb{R}^{L_q \times 1}$ with the dequantization module,\footnote{Note that quantization refers to using a limited number of bits to represent each element of $\mathbf{v}$ with loss of precision, and dequantization refers to the inverse operation of quantization to recover $\mathbf{v}$ from the acquired bit stream\cite{chen2019novel,zhou2023transformerbased}.} and then reconstructs the eigenvector-based CSI matrix with the reconstruction module.

Concretely, the CSI feature vector $\mathbf{v}$ can be expressed by
\begin{equation}\label{tra comp}
\mathbf{v} = F^\prime_\text{com}(\mathbf{W};\mathbf{\Theta}^\prime_\text{com}),
\end{equation}
where $F^\prime_\text{com}$ represents the compression module parameterized by $\mathbf{\Theta}^\prime_\text{com}$. The relation of the recovered CSI feature vector $\hat{\mathbf{v}}$ at the BS and the CSI feature vector $\mathbf{v}$ at the UE can be expressed by
\begin{equation}
\hat{\mathbf{v}} = F^\prime_\text{deq}(F^\prime_\text{quan}(\mathbf{v};\mathbf{\Theta}^\prime_\text{quan});\mathbf{\Theta}^\prime_\text{deq}),
\end{equation}
where $F^\prime_\text{quan}$ and $F^\prime_\text{deq}$ represent the quantization module and the dequantization module parametermized by $\mathbf{\Theta}^\prime_\text{quan}$ and $\mathbf{\Theta}^\prime_\text{deq}$, respectively. The reconstructed eigenvector-based CSI matrix can be expressed by
\begin{equation}\label{tra rec}
\hat{\mathbf{W}} = F^\prime_\text{rec}(\hat{\mathbf{v}};\mathbf{\Theta}^\prime_\text{rec}),
\end{equation}
where $F^\prime_\text{rec}$ represents the reconstruction module parameterized by $\mathbf{\Theta}^\prime_\text{rec}$.

According to the above CSI feedback procedure, traditional deep learning-based methods compress $\mathbf{W}$ directly to obtain $\mathbf{v}$. On the one hand, as the number of antennas increases and the operating bandwidth widens, the dimension of $\mathbf{W}$ expands progressively, leading to rapidly increasing feedback overhead and computational complexity. Although the way that the SampleDL in \cite{wang2021compressive} adopts (i.e., directly downsamples frequency domain subcarriers to achieve a low-dimensional CSI matrix) for CSI feedback is effective in reducing feedback overhead and computational complexity, the downsampling process may lead to severe information loss due to the direct dropping of the majority of subcarriers. On the other hand, training the CSI feedback network with good performance under few-shot scenarios is of great practical significance. Therefore, to simultaneously reduce the feedback overhead and enable the few-shot CSI feedback for massive MIMO systems, we propose the IEFSF. The IEFSF mainly comprises two parts, i.e., the incorporation-extrapolation CSI feedback scheme and the data augmentation methods. In the following section, we first present the proposed incorporation-extrapolation CSI feedback scheme.

\section{Incorporation-Extrapolation CSI Feedback Scheme} \label{Sec. IE}

In this section, we propose the incorporation-extrapolation CSI feedback scheme. Instead of directly compressing the full-dimensional CSI matrix $\mathbf{W}$, a low-dimensional CSI matrix, i.e., $\overline{\mathbf{W}}$, is first formed through the \textbf{incorporation} process at the UE. Then, $\overline{\mathbf{W}}$ is compressed and fed back to the BS. Finally, the acquired $\overline{\mathbf{W}}$ at the BS is \textbf{extrapolated} to recover the full-dimensional CSI matrix $\mathbf{W}$ via the designed frequency extrapolation network.

\subsection{Incorporation Process}

Thanks to the frequency domain correlation among adjacent subcarriers, we can derive a low-dimensional CSI matrix from $\mathbf{W}$ for CSI feedback by incorporating the information of neighboring subcarriers together. Note that, although there is a correlation between adjacent subcarriers, the specific channel information (e.g., angle of departure, channel gain, etc) at different subcarriers may still differ from each other\cite{wang2020block}, especially when the subcarrier spacing is large. Therefore, different from SampleDL which directly drops subcarriers, our proposed incorporation process is designed to merge the channel information of adjacent subcarriers together to try to preserve the channel information as much as possible. Specifically, denoting the set containing the channel matrices of all subcarriers as $\mathcal{H}_{subc} = \{\mathbf{H}_i|i=1,2,\ldots,N_c\}$, we first divide all subcarriers into $N_{grp} = N_c / N_{gr}$ groups, i.e., $\mathcal{H}_{grp}^j$, $j = 1, 2, \ldots, N_{grp}$, with a given granularity $N_{gr}$.\footnote{For simplicity, we assume that $N_c$ is divisible by $N_{gr}$.} Concretely, the $j$-th group $\mathcal{H}_{grp}^j$ can be expressed by
\begin{align} \label{group}
\mathcal{H}_{grp}^j = \{\mathbf{H}_i|i=j*N_{gr}-N_{gr}+1,\ldots,j*N_{gr}\}.
\end{align}

Then, we calculate the Gram matrix $\overline{\mathbf{R}}_j \in \mathbb{C}^{N_T \times N_T}$ of the $j$-th group $\mathcal{H}_{grp}^j$ by averaging the Gram matrices of its $N_{gr}$ constituent subcarriers as
\begin{equation} \label{sb_correlation_mat}
\overline{\mathbf{R}}_j = \frac{1}{N_{gr}} \sum_{i=(j*N_{gr}-N_{gr}+1)}^{j*N_{gr}} \mathbf{H}_i^{H}\mathbf{H}_i.
\end{equation}

Next, the corresponding dominant eigenvector $\overline{\mathbf{w}}_j \in \mathbb{C}^{N_T \times 1}$ for $\overline{\mathbf{R}}_j$ can be calculated as
\begin{equation} \label{sb_eigen}
\overline{\mathbf{R}}_j \overline{\mathbf{w}}_j = \bar{\lambda}^\text{max}_j \overline{\mathbf{w}}_j,
\end{equation}
where $\bar{\lambda}^\text{max}_j$ is the maximum eigenvalue of $\overline{\mathbf{R}}_j$.

To this end, we have the eigenvector-based CSI matrix to be fed back to the BS as
\begin{equation}\label{sb_whole_eigen}
\overline{\mathbf{W}} = [\overline{\mathbf{w}}_1, \overline{\mathbf{w}}_2, \ldots, \overline{\mathbf{w}}_{N_{grp}}].
\end{equation}

Through the incorporation process, i.e., (\ref{group}) - (\ref{sb_whole_eigen}), the information of all subcarriers is incorporated into the CSI matrix $\overline{\mathbf{W}} \in \mathbb{C}^{N_T \times N_{grp}}$. Since $N_{grp} \ll N_c$, the CSI feedback overhead and the computational complexity of the compression module can be significantly reduced when compared to $\mathbf{W} \in \mathbb{C}^{N_T \times N_{c}}$. Specifically, the feedback overhead of $\overline{\mathbf{W}}$ is only $\frac{1}{N_{gr}}$ of that of $\mathbf{W}$.

It is worth emphasizing that the selection of $N_{gr}$ depends on the channel coherent bandwidth. A large channel coherent bandwidth allows a large $N_{gr}$. On the contrary, a small coherent bandwidth allows only a small $N_{gr}$ for recovering the full CSI information from the incorporated partial information. Despite that, according to the incorporation process (i.e., (\ref{group}) - (\ref{sb_whole_eigen})),  the group-level granularity $N_{gr}$ determines the CSI feedback overhead and computational complexity. A large $N_{gr}$ (e.g., $64$) leads to a much small size $\overline{\mathbf{W}}$ and extremely low feedback overhead and computational complexity, while a small $N_{gr}$ (e.g., $2$) can only slightly reduce the dimension of $\overline{\mathbf{W}}$ compared to $\mathbf{W}$ and thereby can not effectively facilitate the CSI feedback overhead and computational complexity reduction. Therefore, a trade-off should be considered in pre-determining an appropriate $N_{gr}$ regarding the channel coherent bandwidth and the CSI feedback overhead and computational complexity.

\subsection{CSI Compression and Reconstruction}
After the incorporation process, similar to (\ref{tra comp}) - (\ref{tra rec}), the compression, quantization, dequantization, and reconstruction operations for $\overline{\mathbf{W}}$ are similar to those for $\mathbf{W}$, which can be specified as
\begin{equation} \label{comp}
\overline{\mathbf{v}} = F_\text{com}(\overline{\mathbf{W}};\mathbf{\Theta}_\text{com}),
\end{equation}
\begin{equation}
\hat{\overline{\mathbf{v}}} = F_\text{deq}(F_\text{quan}(\overline{\mathbf{v}};\mathbf{\Theta}_\text{quan});\mathbf{\Theta}_\text{deq}),
\end{equation}
\begin{equation} \label{rec}
\hat{\overline{\mathbf{W}}} = F_\text{rec}(\hat{\overline{\mathbf{v}}};\mathbf{\Theta}_\text{rec}),
\end{equation}
where $F_\text{com}$ represents the compression module parameterized by $\mathbf{\Theta}_\text{com}$, $F_\text{rec}$ represents the reconstruction module parameterized by $\mathbf{\Theta}_\text{rec}$, and $F_\text{quan}$ and $F_\text{deq}$ represent the quantization module and the dequantization module parametermized by $\mathbf{\Theta}_\text{quan}$ and $\mathbf{\Theta}_\text{deq}$, respectively. Based on (\ref{comp}) - (\ref{rec}), we have
\begin{equation} \label{en-de}
\hat{\overline{\mathbf{W}}} = F_\text{de}(F_\text{en}(\overline{\mathbf{W}};\mathbf{\Theta}_\text{en});\mathbf{\Theta}_\text{de}),
\end{equation}
where $F_\text{en}$ represents the encoder composed of the compression and quantization modules and parameterized by $\mathbf{\Theta}_\text{en} = \mathbf{\Theta}_\text{com} \cup \mathbf{\Theta}_\text{quan}$, and $F_\text{de}$ represents the decoder composed of the dequantization and reconstruction modules and parameterized by $\mathbf{\Theta}_\text{de} = \mathbf{\Theta}_\text{rec} \cup \mathbf{\Theta}_\text{deq}$.

\begin{figure*}[htbp]
\centering
\includegraphics[width=\textwidth]{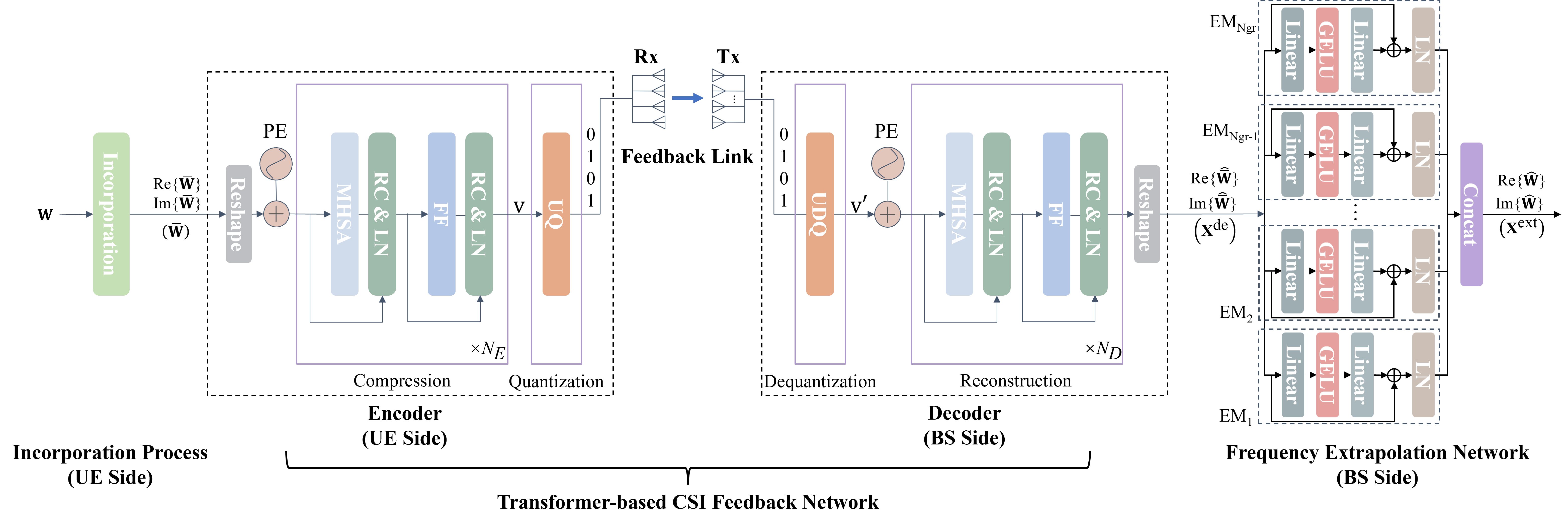}
\caption{The incorporation-extrapolation CSI feedback scheme which consists of the incorporation process, the proposed Transformer-based CSI compression and reconstruction network, and the frequency extrapolation network.}
\label{NN}
\end{figure*}

Note that with an appropriate $N_{gr}$, the frequency domain correlation can be retained at the group level. Therefore, by exploiting the group-level frequency domain correlation, the CSI matrix at adjacent subcarrier groups can be effectively compressed and reconstructed. In the proposed framework, we leverage the Transformer network to explore the correlation among subcarrier groups for CSI feedback. As shown in Fig. \ref{NN}, the compression module and the reconstruction module of the Transformer-based CSI compression and reconstruction network include $N_{com}$ and $N_{rec}$ identical layers, respectively. Each of these identical layers consists of two components: a multi-head self-attention (MHSA) sub-layer and a feed-forward (FF) sub-layer. The MHSA sub-layer performs computations on different subcarrier groups to generate a self-attention matrix. This self-attention matrix plays a crucial role in paying more attention to correlated inputs, thereby improving the effectiveness of the high-dimensional feature learning of the CSI matrix. Such improved feature learning facilitates the subsequent CSI compression and reconstruction operations. The FF sub-layer is designed to enhance the non-linear representation capability of the layer. To further optimize the training process, residual connection (RC)\cite{he2016deep} and layer normalization (LN)\cite{ba2016layer} operations are applied after both the MHSA sub-layer and the FF sub-layer. These operations help improve the neural network's training efficiency. Additionally, since the MHSA sub-layer treats all subcarrier groups equally and does not consider their positional information, supplementary positional encodings (PEs) are inserted before the compression module and the reconstruction module to compensate for this limitation, such that the positional information can also be utilized for better CSI compression and reconstruction. Finally, we exploit uniform quantization (UQ) and corresponding uniform dequantization (UDQ) for conversion between the bit stream and CSI feature vector in the proposed Transformer-based CSI compression and reconstruction network.

\subsection{Frequency Extrapolation Network}

Owing to the frequency domain correlations, the CSI information of all subcarriers, i.e., $\hat{\mathbf{W}} \in \mathbb{C}^{N_T \times N_{c}}$, can be attained through frequency domain extrapolation based on $\hat{\overline{\mathbf{W}}} \in \mathbb{C}^{N_T \times N_{grp}}$. According to (\ref{sb_correlation_mat}) - (\ref{sb_eigen}), the CSI information of $N_{gr}$ subcarriers $\{\mathbf{H}_i|i=j*N_{gr}-N_{gr}+1,j*N_{gr}-N_{gr}+2,\ldots,j*N_{gr}\}$ constituting the $j$-th group $\mathcal{H}_{grp}^j$ is incorporated into $\overline{\mathbf{w}}_j$. Therefore, it is anticipated that $\mathbf{w}_i$ corresponding to $\mathbf{H}_i$ in $\mathcal{H}_{grp}^j$ can be recovered from $\overline{\mathbf{w}}_j$. To recover $\mathbf{w}_i$, depending on the coherent bandwidth and the subcarrier spacing, two kinds of frequency domain correlations can be exploited, i.e., 1) the correlation among subcarrier groups and 2) the correlation among the subcarriers within the same subcarrier group. For ease of elaboration, these two correlations are named 1) the group-level frequency domain correlation and 2) the subcarrier-level frequency domain correlation hereafter.

In order to achieve the full CSI matrix $\hat{\mathbf{W}}$ from $\hat{\overline{\mathbf{W}}}$, the frequency extrapolation network is designed as
\begin{equation}\label{freq. extra.}
\hat{\mathbf{W}} = F_\text{ext}(\hat{\overline{\mathbf{W}}};\mathbf{\Theta}_\text{ext}),
\end{equation}
where $F_\text{ext}$ represents the frequency extrapolation network parameterized by $\mathbf{\Theta}_\text{ext}$. Notice that the principle of the frequency extrapolation in (\ref{freq. extra.}) is different from that of the SampleDL in \cite{wang2021compressive} since the SampleDL generates the low-dimensional CSI matrix by directly dropping most subcarriers, making the recovery of these dropped subcarriers difficult. In addition, the SampleDL first conducts the zero filling interpolation to reconstruct the full-dimensional CSI matrix and then refines it with the refine neural network. This thus requires a refine neural network to recover missing information in a large dimension, which brings significant difficulty in information recovery and also high computational complexity.

{In contrast, to extrapolate $\hat{\overline{\mathbf{W}}}$ to $\hat{\mathbf{W}}$ with low computational complexity, we design a lightweight but effective frequency extrapolation network by exploiting multi-path extrapolation and residual learning based on the subcarrier-level and the group-level frequency domain correlations, as shown in Fig. \ref{NN}. The frequency extrapolation network is composed of $N_{gr}$ extrapolation modules (EMs) and one concatenation module. Each EM further consists of two linear layers, one Gaussian Error Linear Unit (GELU) activation layer, one residual connection operation\cite{he2016deep}, and one layer normalization operation\cite{ba2016layer}. The residual connection is designed to allow the activation from one layer to be directly passed to a later layer and thus bypass the intermediate layers. This avoids the issue of vanishing gradients in EMs and enables the designing of deep neural network architectures with improved model capacity for frequency domain extrapolation.} Specifically, denoting the input to the frequency extrapolation network as $\mathbf{X}^{de} \in \mathbb{R}^{(2 \times N_T) \times N_{grp}}$, $\mathbf{X}^{de}$ is first processed by $EM_i$, $i=1, 2, \ldots, N_{gr}$, as
\begin{align}
\mathbf{X}^{EM_i}(1) &= (\operatorname{GELU}(\mathbf{X}^{de}\mathbf{K}_1^i+\mathbf{b}_1^i))\mathbf{K}_2^i+\mathbf{b}_2^i),\\
\mathbf{X}^{EM_i}(2) &= \mathbf{X}^{de} + \mathbf{X}^{EM_i}(1),\\
\mathbf{X}^{EM_i} &= \frac{\mathbf{X}^{EM_i}(2)-\mathbb{E}_{-1}\{\mathbf{X}^{EM_i}(2)\}}{\sqrt{\operatorname{Var}_{-1}\{\mathbf{X}^{EM_i}(2)\}+\epsilon}} \odot \mathbf{g}_{i} + \mathbf{l}_{i},
\end{align}
where $\mathbf{X}^{EM_i}(1)$, $\mathbf{X}^{EM_i}(2)$, and $\mathbf{X}^{EM_i}$ are the outputs of the GELU activation layer, the residual connection operation, and the layer normalization operation of the $i$-th EM, respectively. $\operatorname{GELU}(\cdot)$ denotes the GELU activation function, $\mathbf{K}_1^i \in \mathbb{R}^{N_{grp} \times N_{grp}}$ and $\mathbf{K}_2^i \in \mathbb{R}^{N_{grp} \times N_{grp}}$ are learnable weight matrices, $\mathbf{b}_1^i \in \mathbb{R}^{1 \times N_{grp}}$ and $\mathbf{b}_2^i \in \mathbb{R}^{1 \times N_{grp}}$ are learnable bias vectors, $\mathbb{E}_{-1}\{\cdot\}$ and $\operatorname{Var}_{-1}\{\cdot\}$ are the expectation and variance of the input matrix along its last dimension, $\odot$ denotes the Hadamard product, $\mathbf{g}_{i} \in \mathbb{R}^{1 \times N_{grp}}$ and $\mathbf{l}_{i} \in \mathbb{R}^{1 \times N_{grp}}$ are learnable affine transformation parameters, and $\epsilon$ is a small real number (e.g. 1e-5) added to the denominator for numerical stability. Then, the outputs of these $N_{gr}$ EMs are concatenated to form the final output $\mathbf{X}^{ext} \in \mathbb{R}^{(2 \times N_T) \times N_{c}}$:
\begin{align}
\mathbf{X}^{ext} = [\mathbf{X}^{EM_1}, \mathbf{X}^{EM_2}, \ldots, \mathbf{X}^{EM_{N_{gr}}}].
\end{align}
Hence, $\hat{\mathbf{W}}$ can be achieved from $\mathbf{X}^{ext}$ by first reshaping $\mathbf{X}^{ext}$ to $\underline{\mathbf{X}}^{ext} \in \mathbb{R}^{2 \times N_T \times N_{c}}$ and then splitting $\underline{\mathbf{X}}^{ext}$ along its first dimension to obtain the real part and the imaginary part of $\hat{\mathbf{W}}$.

{It is worth emphasizing that in the frequency extrapolation network, each EM mainly exploits the group-level frequency domain correlation resorting to residual learning while the multi-path learning architecture mainly exploits the subcarrier-level frequency domain correlation to refine the concatenated outputs of all EMs. And we also emphasize that the frequency extrapolation network is effective but lightweight for the BS side to handle the extrapolation process. This is because the proposed frequency extrapolation network performs frequency domain extrapolation via multi-path learning and thus avoids direct extrapolation at the subcarrier level and achieves significant computational efficiency compared with existing convolutional extrapolation networks\cite{dong2019deep, dong2019deepa, mashhadi2021pruning}. The computational complexity of the proposed frequency extrapolation network (FEN) and the convolutional extrapolation network (CEN) in \cite{dong2019deep, dong2019deepa, mashhadi2021pruning} can be expressed by\footnote{Note that we also consider a two-layer architecture similar to our frequency extrapolation network for the convolutional extrapolation network for a fair comparison.}
\begin{align}
\mathcal{O}(FEN) &= \mathcal{O}(4 N_{gr} N_T N_{grp}^2) = \mathcal{O}(4 N_T N_{grp} N_c), \label{complexity FEN}\\
\mathcal{O}(CEN) &= \mathcal{O}(2 K N_T N_{grp} N_{c} + 2 K N_T N_{c}^2), \label{complexity CEN}
\end{align}
where $K>=1$ is the kernel size of the convolutional extrapolation network. Referring to (\ref{complexity FEN}) and (\ref{complexity CEN}), the proposed extrapolation network is much more computationally efficient than the convolutional extrapolation network since $N_{grp} \ll N_c$.}

\subsection{Model Training}
In the model training stage, the proposed Transformer-based CSI compression and reconstruction network and the proposed frequency extrapolation network are trained jointly in an end-to-end manner. The loss function used for model training is defined as
\begin{equation}
L(\mathbf{W}, \hat{\mathbf{W}}) = 1 - \frac{1}{N_{c}} \sum_{i=1}^{N_{c}} \rho_i^2,
\end{equation}
where $\rho_i$ is the generalized cosine similarity (GCS) of the $i$-th subcarrier defined as
\begin{equation}
\rho_i = \frac{\| \mathbf{w}_i^{H} \hat{\mathbf{w}}_i \|_2}{\| \mathbf{w}_i \|_2 \|  \hat{\mathbf{w}}_i \|_2 },
\end{equation}
where $\|\cdot\|_2$ denotes the L2 norm.

The proposed two networks are jointly trained by minimizing the loss function with the Adam optimizer \cite{kingma2014adam}:
\begin{align} \label{loss}
(\mathbf{\Theta}_{de}^*, &\mathbf{\Theta}_{en}^*, \mathbf{\Theta}_{ext}^*) = \mathop{\arg\min}_{\mathbf{\Theta}_{en},\mathbf{\Theta}_{de},\mathbf{\Theta}_{ext}} \nonumber \\
& \mathbb{E}\{ L\left(\mathbf{W}, F_{ext}(F_{de}(F_{en}(\overline{\mathbf{W}}; \mathbf{\Theta}_{en}); \mathbf{\Theta}_{de}); \mathbf{\Theta}_{ext})\right)\},
\end{align}
where $\mathbb{E}(\cdot)$ is the expectation over a batch of training samples, $F_{en}$ and $F_{de}$ represent the encoder and decoder of the Transformer-based CSI compression and reconstruction network with learnable parameters $\mathbf{\Theta}_{en}$ and $\mathbf{\Theta}_{de}$, respectively, and $F_{ext}$ represents the frequency extrapolation network with learnable parameters $\mathbf{\Theta}_{ext}$.

\section{Data Augmentation Methods for Few-Shot Learning} \label{Sec. DataAug}
To fulfill few-shot CSI feedback and reduce large amounts of collected samples for model training, we propose the knowledge-driven data augmentation method and the AIGC-based data augmentation method in this section. By combining the knowledge-driven data augmentation method with the AIGC-based data augmentation method, a comprehensive augmented dataset with both sufficient data samples and high sample diversity can be obtained for few-shot CSI feedback.

\subsection{Knowledge-Driven Data Augmentation Method}
\begin{figure}[htbp]
\centering
\includegraphics[width=0.5\textwidth]{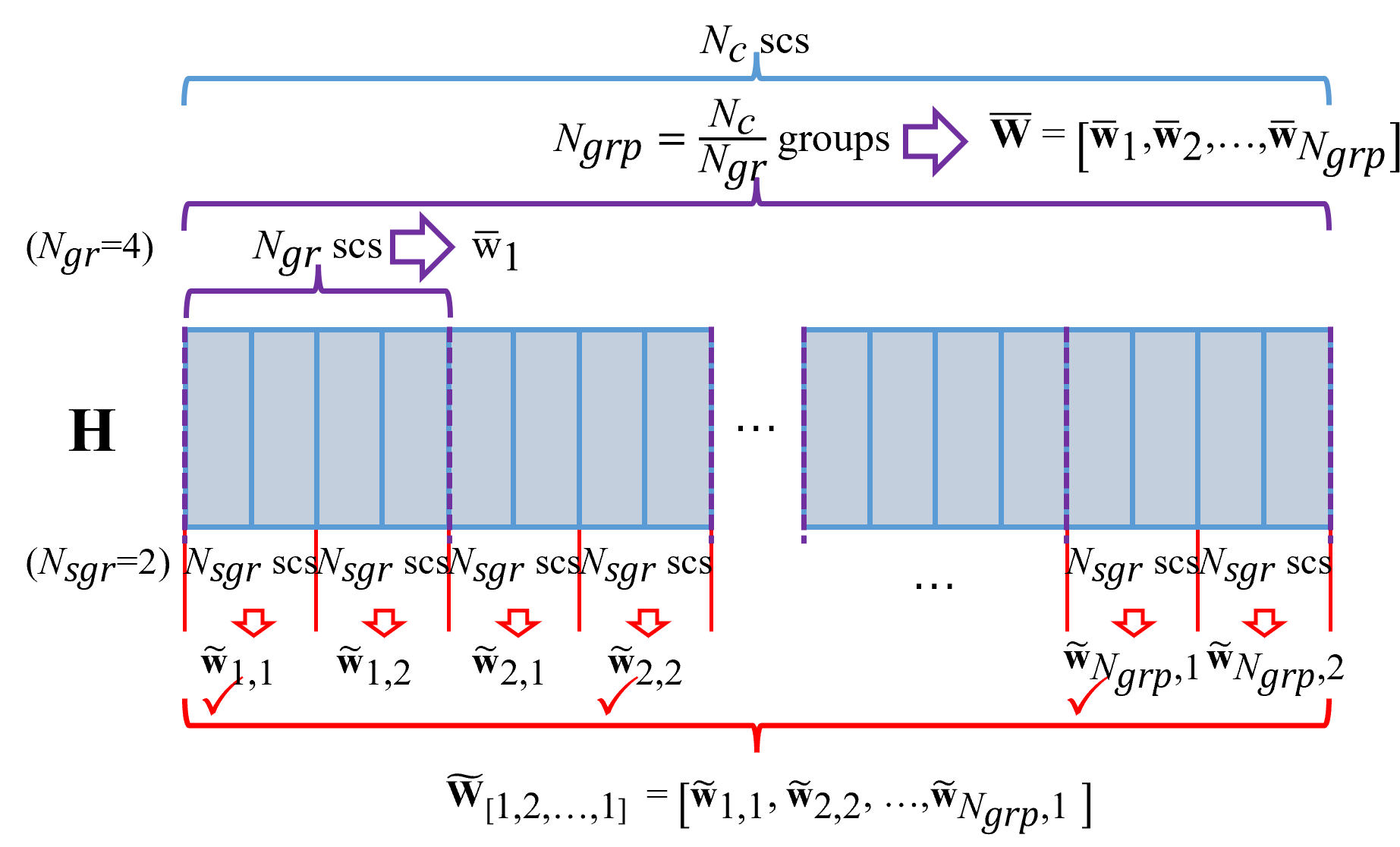}
\caption{An illustration of the knowledge-driven data augmentation method ($N_{gr}=4$ and $N_{sgr}=2$). scs: subcarriers.}
\label{kdda}
\end{figure}

According to the channel property, there are both subcarrier-level and group-level frequency domain correlations that can be leveraged to augment data samples. Based on this domain knowledge, we propose the knowledge-driven data augmentation method for CSI feedback. Recall that in the incorporation process, every $N_{gr}$ subcarriers in $\mathcal{H}_{subc}$ are grouped together and each group $\mathcal{H}_{grp}^j$ yields one dominant eigenvector $\overline{\mathbf{w}}_j$. Intuitively, $\overline{\mathbf{W}} = [\overline{\mathbf{w}}_1, \overline{\mathbf{w}}_2, \ldots, \overline{\mathbf{w}}_{N_{grp}}]$ corresponding to all groups $\{\mathcal{H}_{grp}^j|j=1,2,\ldots,N_{grp}\}$ from collected samples can be directly used for training the CSI feedback network. To augment the training samples by fully leveraging the frequency domain correlations, we also consider a small granularity $N_{sgr} < N_{gr}$.\footnote{To keep highly correlated frequency domain correlations in the augmented samples, we only consider a small granularity, i.e., $N_{sgr} < N_{gr}$.} As a result, the $j$-th group, i.e., $\mathcal{H}_{grp}^j$, can be further divided into $N_{sgrp} = N_{gr}/N_{sgr}$ subgroups, i.e., $\mathcal{H}_{sgrp}^{j,k}$, $k = 1, 2, \ldots, N_{sgrp}$. And we have
\begin{align} \label{subgroup}
\mathcal{H}_{sgrp}^{j,k} &= \{\mathbf{H}_i|i=j*N_{gr} + k*N_{sgr} - N_{gr} - N_{sgr} + 1,\\ \notag
&\ldots,j*N_{gr} + k*N_{sgr} - N_{gr}\}.
\end{align}
Then, similar to (\ref{sb_correlation_mat}) - (\ref{sb_eigen}), we obtain dominant eigenvectors for the $N_{sgrp}$ subgroups $\{\mathcal{H}_{sgrp}^{j,1}, \mathcal{H}_{sgrp}^{j,2}, \ldots, \mathcal{H}_{sgrp}^{j,N_{sgrp}}\}$ of the $j$-th group $\mathcal{H}_{grp}^j$ as

\begin{equation} \label{ssb_correlation_mat}
\widetilde{\mathbf{R}}_{j,k} = \frac{1}{N_{sgr}} \sum_{i=(j*N_{gr} + k*N_{sgr} - N_{gr} - N_{sgr} + 1)}^{j*N_{gr} + k*N_{sgr} - N_{gr}} \mathbf{H}_i^{H}\mathbf{H}_i,
\end{equation}
\begin{equation} \label{ssb_eigen}
\widetilde{\mathbf{R}}_{j,k} \widetilde{\mathbf{w}}_{j,k} = \bar{\lambda}^\text{max}_{j,k} \widetilde{\mathbf{w}}_{j,k}.
\end{equation}

Hence, by collecting all subgroups' dominant eigenvectors for each group, we form an augmented sample of $\overline{\mathbf{W}}$ as

\begin{equation}\label{ssb_whole_eigen}
\widetilde{\mathbf{W}}_\mathbf{m} = [\widetilde{\mathbf{w}}_{1,m_1}, \widetilde{\mathbf{w}}_{2,m_2}, \ldots, \widetilde{\mathbf{w}}_{N_{grp},m_{N_{grp}}}],
\end{equation}
where $\mathbf{m}=[m_1, m_2, \ldots, m_{N_{grp}}]$ and $m_n \in \{1,2,\ldots,N_{sgrp}\}$ for $n = 1,2,\ldots,N_{grp}$. Therefore, theoretically, with different settings of $\mathbf{m}$, ${(C_{N_{sgrp}}^1)}^{N_{grp}}$ (here $C_a^b$ denotes the combination operator indicating choosing $b$ elements from $a$ elements) augmented samples can be generated for $\overline{\mathbf{W}}$.\footnote{It is worth mentioning that we can also group non-continuous subcarriers together to further increase the number of augmented samples. However, grouping non-continuous subcarriers together would increase the computational complexity of the KDDA method but generate similar augmented samples, which is not very efficient.} For ease of understanding, Fig. \ref{kdda} gives an illustration of the knowledge-driven data augmentation method with $N_{gr}=4$ and $N_{sgr}=2$, where $N_{grp}$ eigenvectors (marked with red ticks) from subgroups are selected to form an augmented eigenvector-based CSI matrix $\widetilde{\mathbf{W}}_{[1,2,\ldots,1]}$ with $\mathbf{m}=[1,2,\ldots,1]$.

Note that the proposed knowledge-driven data augmentation method can be easily implemented to yield extensive augmented samples, which is very efficient in improving the performance of CSI feedback, especially under extremely few-shot CSI feedback scenarios. Nevertheless, such augmented samples are generated relying on frequency domain correlations from a limited number of practically collected samples, which thus lack of sample diversity. To further improve the diversity of augmented samples, we resort to AIGC-based data augmentation methods and propose the EGAN to generate diverse samples by learning the distribution of data samples.

\subsection{AIGC-based Data Augmentation Method}

\begin{figure}[htbp] 
\centering
\includegraphics[width=0.5\textwidth]{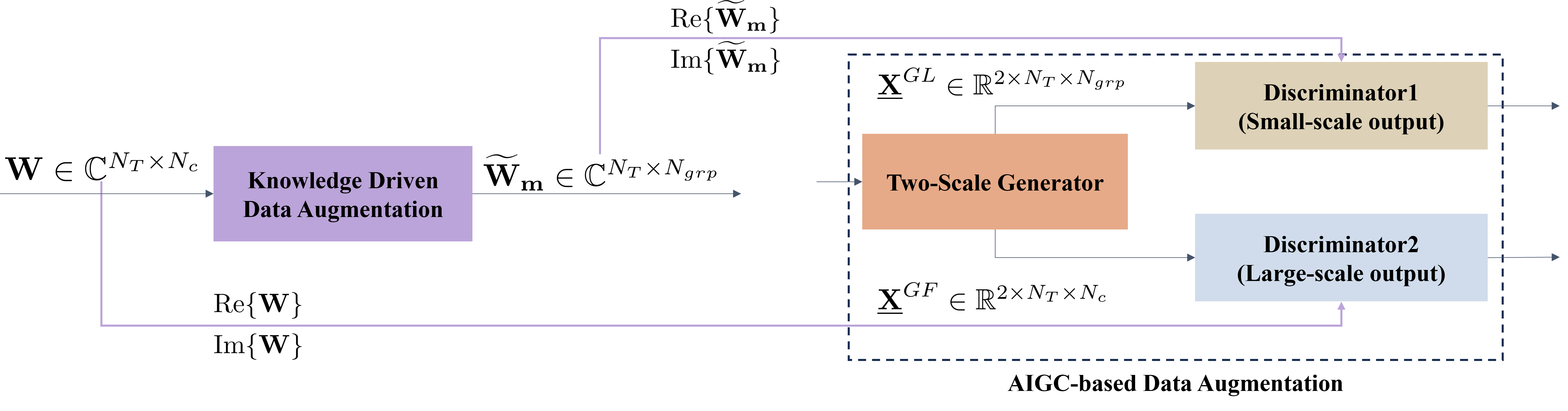}
\caption{The diagram depicting the connection between the knowledge-driven data augmentation method and the AIGC-based data augmentation method.}
\label{kdda-aigc}
\end{figure}

Existing AIGC-based data augmentation methods for few-shot CSI feedback augment data samples by first learning the distribution of collected data samples and then generating diverse samples following the learned distribution via generative models (e.g., GANs). However, compared with discriminative models, generative models typically require large amounts of training data\cite{li2020fewshot} to model the distribution well, which contradicts the intention of exploiting AIGC for data augmentation under few-shot scenarios. {To solve this issue, we propose a novel AIGC-based data augmentation method, which is able to model the distribution of the CSI matrices well with a few collected samples. Specifically, a new GAN architecture, called EGAN, is proposed to learn the distributions of CSI matrices on two scales, i.e., the low-dimensional CSI matrices $\overline{\mathbf{W}}$ and the full-dimensional CSI matrices $\mathbf{W}$. Compared with vanilla GANs, the EGAN benefits from learning an additional distribution of low-dimensional CSI matrices, given that low-dimensional CSI matrices contain internal information of full-dimensional CSI matrices thanks to the incorporation process. This is beneficial for the generation of full-dimensional CSI matrices. Moreover, as depicted in Fig. \ref{kdda-aigc}, the knowledge-driven data augmentation method is also adopted to generate augmented data samples for training the EGAN with sufficient samples under few-shot scenarios.}

\subsubsection{Network Architecture of the EGAN}

\begin{figure}[htbp] 
\centering
\includegraphics[width=0.5\textwidth]{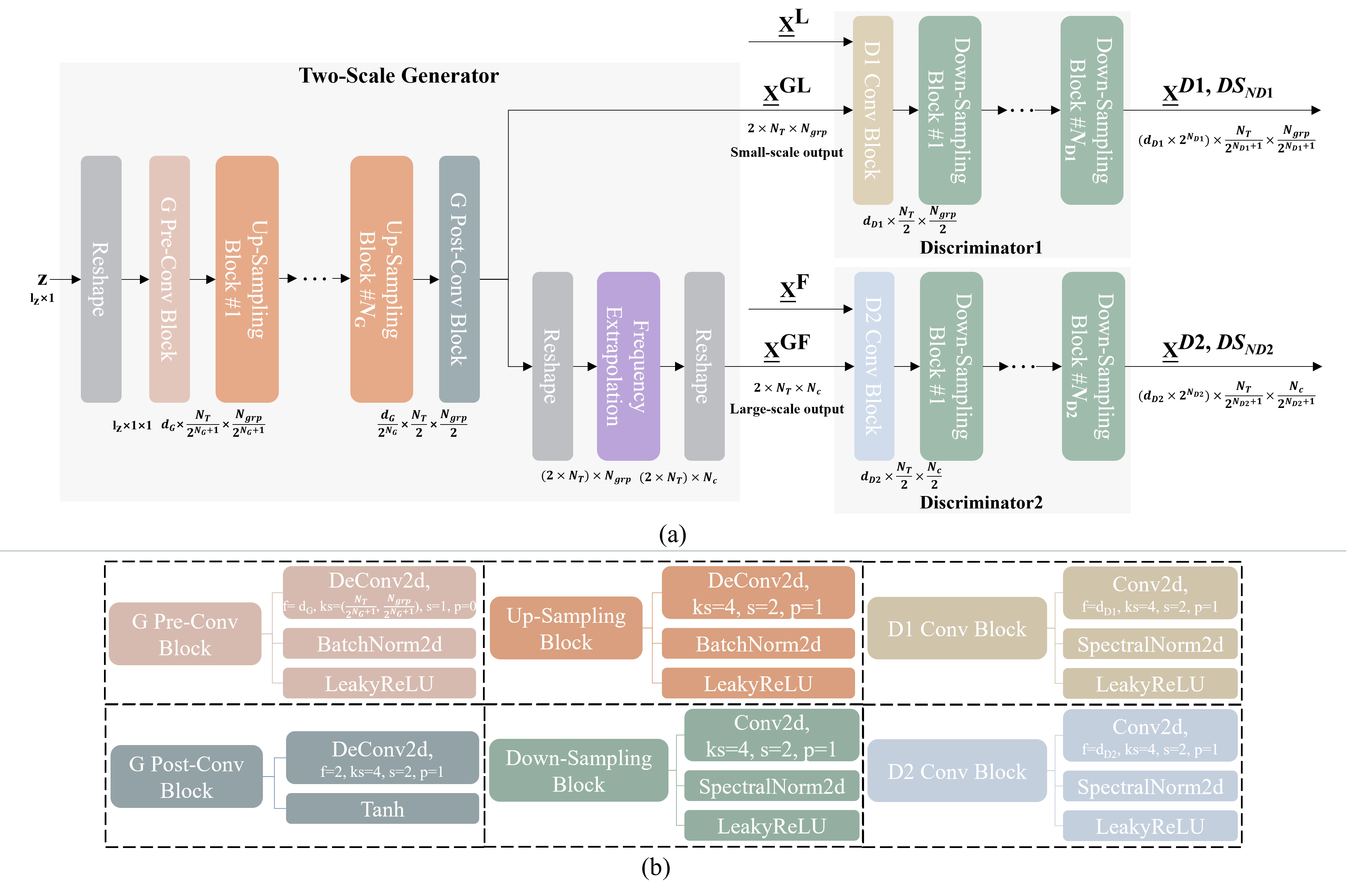}
\caption{The network architecture of the EGAN. (a) The overall network architecture of the EGAN. The sizes of key tensors and neural block outputs are given near them. (b) Detailed architectures and configurations of neural blocks constituting the EGAN. The detailed architecture and configuration of the frequency extrapolation network have been given in Fig. \ref{NN} and are therefore neglected here.}
\label{egan}
\end{figure}

The network architecture of the EGAN is shown in Fig. \ref{egan}. Inspired by \cite{shaham2019singan}, we design the EGAN with a two-scale generator and two associated discriminators. The two-scale generator generates a low-dimensional CSI matrix (small-scale output) and a full-dimensional CSI matrix (large-scale output) at the same time, and the two discriminators (marked by `Discriminator1' and `Discriminator2') distinguish low-dimensional/full-dimensional generated samples from low-dimensional/full-dimensional real samples. The EGAN learns both the distribution of full-dimensional CSI matrices and the distribution of low-dimensional CSI matrices. Since real low-dimensional CSI matrices are obtained with the incorporation process, the information of full-dimensional CSI matrices is implicitly contained in these low-dimensional CSI matrices. Therefore, learning from both full-dimensional CSI matrices and low-dimensional CSI matrices will enhance the generation of full-dimensional CSI matrices and also will help in training generative models under few-shot scenarios.

The input to the generator, i.e., the noise vector $\mathbf{z} \in \mathbb{R}^{l_z \times 1}$, is first rearranged and then pre-processed by the pre-conv block, which yields tensor $\underline{\mathbf{X}}^{{US}_0} \in \mathbb{R}^{d_G \times \frac{N_T}{2^{N_G+1}} \times \frac{N_{grp}}{2^{N_G+1}}}$ as
\begin{equation}
\underline{\mathbf{X}}^{{US}_0} = \operatorname{LR}(\operatorname{BN}(\operatorname{DeConv}^{\frac{N_T}{2^{N_G+1}} \times \frac{N_{grp}}{2^{N_G+1}}}_{1,0}(\mathrm{R}(\mathbf{z})))),
\end{equation}
where $d_G$ is a hyper-parameter as the initial feature dimension of the generator; $\operatorname{LR}(\cdot)$ and $\operatorname{BN}(\cdot)$ denote the leaky rectified linear unit activation function and batch normalization operation\cite{ioffe2015batch}, respectively; $\operatorname{DeConv}^{a \times b}_{s,p}(\cdot)$ denotes the 2-dimensional deconvolution operation with the kernel size being $a \times b$, stride being $s$, and padding size being $p$; $\mathrm{R}(\cdot)$ denotes the rearrangement operation, and $N_G$ is the number of up-sampling blocks. The $N_G$ up-sampling blocks transform $\underline{\mathbf{X}}^{{US}_0}$ to $\underline{\mathbf{X}}^{{US}_{N_G}} \in \mathbb{R}^{\frac{d_G}{2^{N_G}} \times \frac{N_T}{2} \times \frac{N_{grp}}{2}}$ block-by-block and each block performs $2 \times 2$ upsampling on the output feature map of the previous block, which can be expressed as
\begin{align}
    \underline{\mathbf{X}}^{{US}_i} & \in \mathbb{R}^{\frac{d_G}{2^{i}} \times \frac{N_T}{2^{N_G-i+1}} \times \frac{N_{grp}}{2^{N_G-i+1}}} \nonumber \\
    &= \operatorname{LR}(\operatorname{BN}(\operatorname{DeConv}^{4 \times 4}_{2,1}(\underline{\mathbf{X}}^{{US}_{i-1}}))), i = 1, \ldots, N_G ,
\end{align}
and $N_G$ is determined by
\begin{equation}
    N_G = \operatorname{min}(\log_2^{N_T} - 1, \log_2^{N_{grp}} - 1).
\end{equation}

Then, $\underline{\mathbf{X}}^{{US}_{N_G}}$ is post-processed by the post-conv block, which yields $\underline{\mathbf{X}}^{GL} \in \mathbb{R}^{2 \times N_T \times N_{grp}}$, i.e., the low-dimensional output of the generator, as
\begin{equation}
\underline{\mathbf{X}}^{GL} = \operatorname{Tanh}(\operatorname{DeConv}^{4 \times 4}_{2,1}(\underline{\mathbf{X}}^{{US}_{N_G}})),
\end{equation}
where $\operatorname{Tanh}(\cdot)$ denotes the hyperbolic tangent activation function.

Another output of the generator, i.e., the generated full-dimensional CSI matrix $\underline{\mathbf{X}}^{GF} \in \mathbb{R}^{2 \times N_T \times N_{c}}$, is further generated from $\underline{\mathbf{X}}^{GL}$ with a frequency extrapolation network, which can be specified as
\begin{equation}
\underline{\mathbf{X}}^{GF} = \mathrm{R}(F_{ext}(\mathrm{R}(\underline{\mathbf{X}}^{GL}))).
\end{equation}

Two discriminators are designed to distinguish generated low-dimensional / full-dimensional samples generated by the generator from real low-dimensional / full-dimensional samples, respectively. Each discriminator consists of one convolutional block and several down-sampling blocks. The outputs of these two convolutional blocks are given by
\begin{align}
    \underline{\mathbf{X}}^{{D1,DS}_0} \in \mathbb{R}^{d_{D1} \times \frac{N_T}{2} \times \frac{N_{grp}}{2}} &= \operatorname{LR}(\operatorname{SN}(\operatorname{Conv}^{4 \times 4}_{2,1}(\underline{\mathbf{X}}^{{D1,in}})), \\
    \underline{\mathbf{X}}^{{D2,DS}_0} \in \mathbb{R}^{d_{D2} \times \frac{N_T}{2} \times \frac{N_{c}}{2}} &= \operatorname{LR}(\operatorname{SN}(\operatorname{Conv}^{4 \times 4}_{2,1}(\underline{\mathbf{X}}^{{D2,in}})),
\end{align}
where $d_{D1}$ and $d_{D2}$ are hyper-parameters as the initial feature dimensions of the two discriminators respectively, and $\operatorname{SN}(\cdot)$ denotes the spectral normalization operation\cite{miyato2018spectral}. $\operatorname{Conv}^{a \times b}_{s,p}(\cdot)$ denotes the 2-dimensional convolution operation with the kernel size being $a \times b$, stride being $s$, and padding size being $p$. And the outputs of the down-sampling blocks are given by
\begin{align}
\underline{\mathbf{X}}^{{D1,DS_i}} & \in \mathbb{R}^{(d_{D1} \times 2^{i}) \times \frac{N_T}{2^{i+1}} \times \frac{N_{grp}}{2^{i+1}}} \nonumber \\
&= \operatorname{LR}(\operatorname{SN}(\operatorname{Conv}^{4 \times 4}_{2,1}(\underline{\mathbf{X}}^{{D1,DS}_{i-1}}))), i = 1, \ldots, N_{D1} ,\\
\underline{\mathbf{X}}^{{D2,DS_i}} & \in \mathbb{R}^{(d_{D2} \times 2^{i}) \times \frac{N_T}{2^{i+1}} \times \frac{N_{c}}{2^{i+1}}} \nonumber \\
&= \operatorname{LR}(\operatorname{SN}(\operatorname{Conv}^{4 \times 4}_{2,1}(\underline{\mathbf{X}}^{{D2,DS}_{i-1}}))), i = 1, \ldots, N_{D2} ,
\end{align}
where $N_{D1}$ and $N_{D2}$ are determined by
\begin{align}
    N_{D1} &= \operatorname{min}(\log_2^{N_T} - 1, \log_2^{N_{grp}} - 1), \\
    N_{D2} &= \operatorname{min}(\log_2^{N_T} - 1, \log_2^{N_{c}} - 1).
\end{align}

{Note that the outputs of the two discriminators, i.e., $\underline{\mathbf{X}}^{{D1,DS_{N_{D1}}}}$ and $\underline{\mathbf{X}}^{{D2,DS_{N_{D2}}}}$, are three-dimensional tensors as shown in Fig. \ref{egan}, which can be regarded as distributed representations for the distributions of low-dimensional and full-dimensional samples, respectively. Compared with the vanilla GAN which outputs a probability indicating whether the input is a generated sample or a real sample, a distributed representation in a tensor form keeps more information about the sample distribution and is thus more effective\cite{mikolov2013distributed}.}

\subsubsection{Training of EGAN}
The objective of the generator of EGAN is to generate as real and diverse samples as the collected real samples as possible, while the discriminators of EGAN are responsible for distinguishing generated samples from real samples. To balance the quality and diversity of generated samples, we use the Wasserstein-1 distance\cite{arjovsky2017wasserstein} to measure the distance between the distribution of real samples and the distribution of generated samples as
\begin{equation} \label{vanilla_wd}
W(p_r, p_g) = \inf_{\gamma\in \Pi(p_r,p_g)} \mathbb{E}_{(\underline{\mathbf{X}}_{r}, \underline{\mathbf{X}}_{g}) \sim \gamma}\{\|\underline{\mathbf{X}}_{r} - \underline{\mathbf{X}}_{g}\|_2\},
\end{equation}
where $\inf (\cdot)$ denotes the infimum, $\underline{\mathbf{X}}_{r} \sim p_r$ and $\underline{\mathbf{X}}_{g} \sim p_g$ denote real and generated samples, respectively, and $p_r$ and $p_g$ denote the distributions of real and generated samples, respectively. $\Pi(p_r,p_g)$ denotes the set of all joint distributions $\gamma(\underline{\mathbf{X}}_{r}, \underline{\mathbf{X}}_{g})$ with marginals being $p_r$ and $p_g$. Using the Kantorovich-Rubinstein duality\cite{arjovsky2017wasserstein,villani2009optimal}, (\ref{vanilla_wd}) can be rewritten as
\begin{equation} \label{mod_wd}
W(p_r, p_g) = \sup_{\|f\|_2 \le 1} \mathbb{E}_{\underline{\mathbf{X}}_{r} \sim p_r}\{f(\underline{\mathbf{X}}_{r})\}  - \mathbb{E}_{\underline{\mathbf{X}}_{g} \sim p_g}\{f(\underline{\mathbf{X}}_{g})\},
\end{equation}
where $\sup (\cdot)$ denotes the supremum, and $\|f\|_2 \le 1$ indicates that $f$ is a 1-Lipschitz function.

To stabilize the training of EGAN, two gradient penalty terms \cite{gulrajani2017improved} are exploited to enforce the 1-Lipschitz constraint (i.e., $\|f\|_2 \le 1$) for the two discriminators. Finally, denoting the generator, `Discriminator1', and `Discriminator2' as $G$, $D_1$, and $D_2$, respectively, and denoting the real samples aligned with $\underline{\mathbf{X}}^{GL} \sim p^{GL}$ and $\underline{\mathbf{X}}^{GF} \sim p^{GF}$ as $\underline{\mathbf{X}}^{L} \sim p^L$ and $\underline{\mathbf{X}}^{F} \sim p^F$, respectively, the training of EGAN can be formulated with
\begin{align} \label{egan_loss} 
\min_G \max_{D_1, D_2} \left(\mathbb{E}_{\underline{\mathbf{X}}^L \sim p^L}\{D_1(\underline{\mathbf{X}}^L)\} - \mathbb{E}_{\underline{\mathbf{X}}^{GL} \sim p^{GL}} \{D_1(\underline{\mathbf{X}}^{GL})\}\right. \nonumber \\
+ \mathbb{E}_{\underline{\mathbf{X}}^F \sim p^F}\{D_2(\underline{\mathbf{X}}^F)\} - \mathbb{E}_{\underline{\mathbf{X}}^{GF} \sim p^{GF}} \{D_2(\underline{\mathbf{X}}^{GF})\} \nonumber \\
+ \lambda_1 \mathbb{E}_{\tilde{\underline{\mathbf{X}}}^{L} \sim \tilde{p}^L} \{(\|\nabla_{\tilde{\underline{\mathbf{X}}}^{L}} D_1(\tilde{\underline{\mathbf{X}}}^{L}) \|_2 -1)^2\} \nonumber \\
\left. + \lambda_2 \mathbb{E}_{\tilde{\underline{\mathbf{X}}}^{F} \sim \tilde{p}^F} \{(\|\nabla_{\tilde{\underline{\mathbf{X}}}^{F}} D_2(\tilde{\underline{\mathbf{X}}}^{F}) \|_2 -1)^2\}\right),
\end{align}
where $\tilde{\underline{\mathbf{X}}}^{L} \sim \tilde{p}^L =\alpha \underline{\mathbf{X}}^{L}+(1-\alpha) \underline{\mathbf{X}}^{GL}$ and $\tilde{\underline{\mathbf{X}}}^{F} \sim \tilde{p}^F =\alpha \underline{\mathbf{X}}^{F}+(1-\alpha) \underline{\mathbf{X}}^{GF}$ are randomly interpolated samples with $\alpha \sim U[0,1)$ for gradient penalty, $U$ denotes the uniform distribution, and $\lambda_1$ and $\lambda_2$ are constant penalty coefficients.

{Note that the training of EGAN is a min-max optimization problem and is solved by alternately optimizing the generator and two discriminators. Specifically, the maximization problem (i.e., optimizing the two discriminators while the generator is fixed) and the minimization problem (i.e., optimizing the generator while the discriminators are fixed) are solved alternatively in an adversarial training manner. The discriminators are responsible for distinguishing real and generated samples and thus are optimized to maximize the Wasserstein distance, and the generator is optimized to minimize the Wasserstein distance for generating samples that are more real from the distribution perspective. Therefore, through the adversarial training, the Wasserstein distance will decrease gradually and finally, the generator will generate samples following a distribution similar to that of real samples.}{Additionally, the knowledge-driven data augmentation method is adopted to generate augmented data samples for training the EGAN as depicted in Fig. \ref{kdda-aigc}, such that the EGAN can be trained with few collected samples.}

\section{Simulation Results} \label{Sec. Exp.}
In this section, we first introduce the experimental settings and the performance evaluation metrics. Then, we demonstrate the performance of the proposed two data augmentation methods in enabling CSI feedback under few-shot scenarios. {After that, we compare the performance of the proposed CSI feedback framework with existing schemes under different numbers of collected samples, feedback bits, and signal-to-noise ratios (SNRs).} Finally, we analyze the computational complexity of the proposed CSI feedback framework.

\subsection{Simulation Setup}
\subsubsection{Experimental Datasets}

\begin{figure}[htbp]
\centering
\subfigure[Indoor scenario (`I3')]{
\begin{minipage}[t]{0.22\textwidth}
\centering
\includegraphics[width=\textwidth]{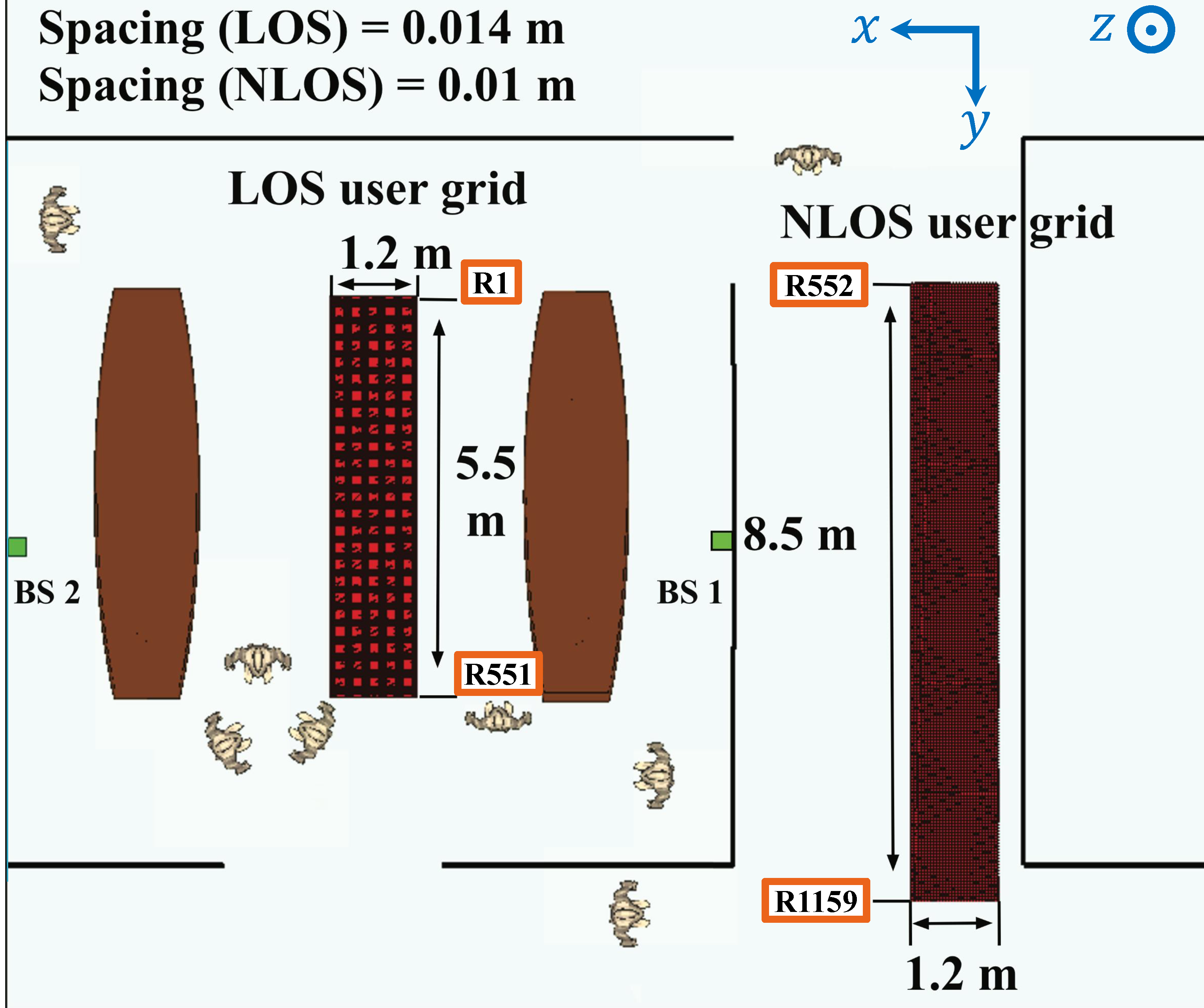}
\end{minipage}
}
\hfill
\subfigure[Outdoor scenario (`O1')]{
\begin{minipage}[t]{0.22\textwidth}
\centering
\includegraphics[width=\textwidth]{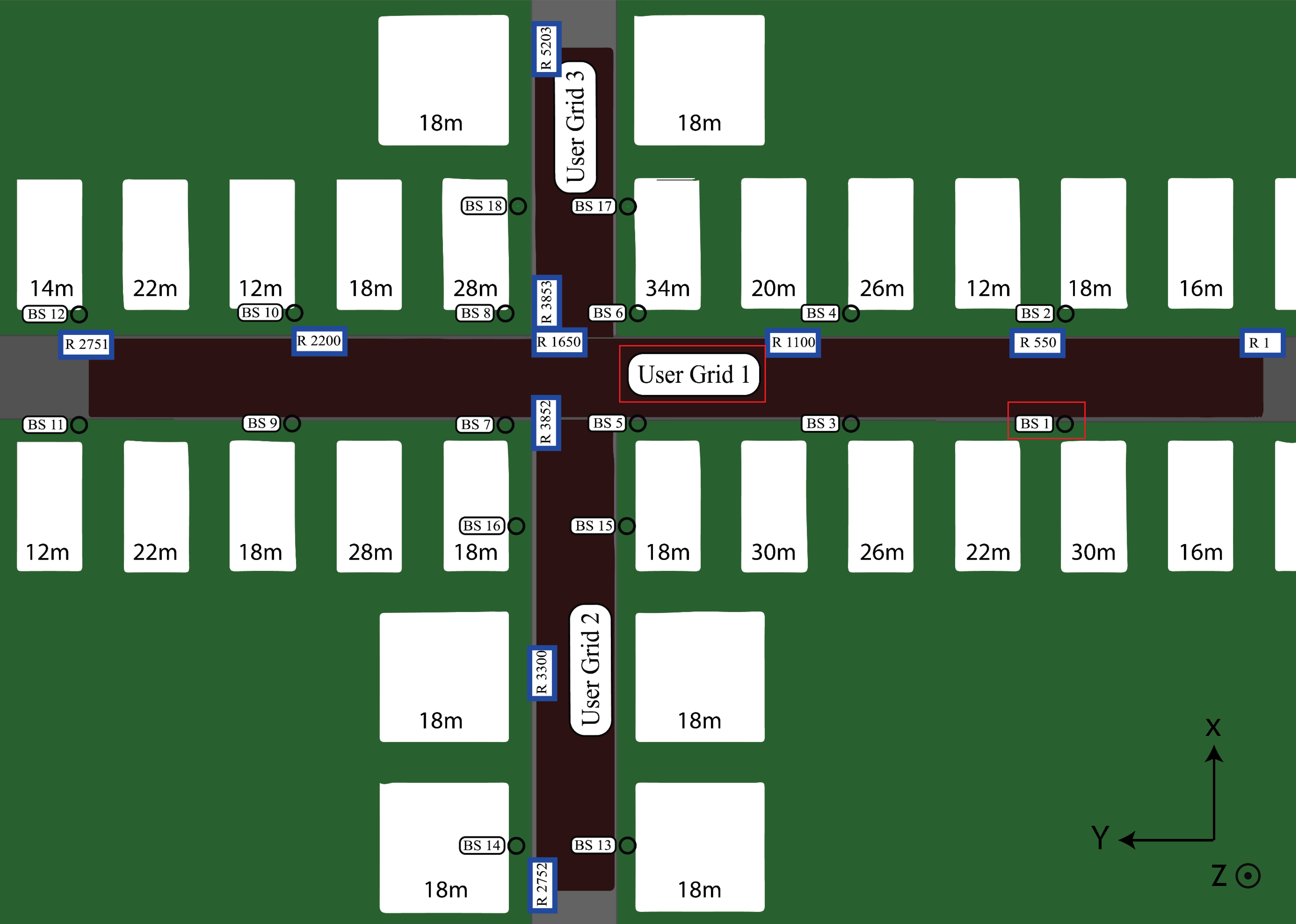}
\end{minipage}
}
\caption{Top views of the indoor scenario and the outdoor scenario.}
\label{TopViews}
\end{figure}

A generic ray tracing-based dataset, i.e., DeepMIMO \cite{alkhateeb2019deepmimo}, is leveraged in the simulations. In the DeepMIMO dataset, the wireless channels are constructed based on ray tracing from the Wireless InSite ray-tracing simulator\cite{wireless_insite}. We consider two scenarios in the dataset, i.e., the indoor scenario (`I3') and the outdoor scenario (`O1'), to validate our proposed CSI feedback framework. The top views and the system setup of the two scenarios are shown in Fig. \ref{TopViews} and Table \ref{Sys. Set.}, respectively. 

The `I3' indoor scenario is an indoor conference room scenario whose 3-dimensional size is $10$ m $\times$ $11$ m $\times$ $3$ m (width $\times$ length $\times$ height), as shown in Fig. \ref{TopViews} (a). Only the `BS1' located at a height of $2$m is active in the experiments, and the channels from the `BS1' to UE locations in the `LOS user grid' are collected for experiments. The system operation frequency band and bandwidth are $60$ GHz and $100$ MHz, respectively. As shown in Fig. \ref{TopViews} (b), the `O1' outdoor scenario is an urban outdoor scenario consisting of two streets and one intersection. The horizontal street is $600$m long and $40$m wide, and the cross street is $440$m long and $40$m wide. The two streets have buildings on both sides, and the heights of these buildings are shown at the top view of the scenario. Also, only the `BS1' located at a height of $6$m is active in the experiments, and the channels from the `BS1' to UE locations in the `User Grid 1' are collected for experiments. The system operation frequency band and bandwidth are $3.5$ GHz and $100$ MHz, respectively.

In both scenarios, the BS is equipped with a uniform linear array (ULA) with $32$ antennas and the UE is equipped with a ULA with $4$ antennas. The systems both operate with OFDM modulation with $1024$ subcarriers. As a result, each channel sample is with the size of $4\times32\times1024$ and the corresponding full eigenvector-based CSI matrix is with the size of $32\times1024$. 

The collected channel datasets for the indoor and outdoor scenarios have $5,000$ and $10,000$ channel samples, respectively. For the indoor scenario, we use only $50$ - $450$ samples for training in the few-shot experiments, while all experiments are validated with $50$ samples and tested with $4,500$ samples. For the outdoor scenario, we use only $100$ - $900$ samples for training, while all experiments are validated with $100$ samples and tested with $9,000$ samples.

{The proposed incorporation-extrapolation CSI feedback scheme and the knowledge-driven data augmentation method essentially rely on frequency domain correlations. To quantify the frequency domain correlation among the the $i$-th subcarrier group and $(i+1)$-th subcarrier group of the $j$-th CSI sample given a group granularity $N_{gr}$, the adjacent GCS, denoted by $\tilde{\rho}_j(i,i+1)$, can be exploited, which is defined as
\begin{equation}
\tilde{\rho}_j(i,i+1) = \frac{\| \mathbf{w}_i^{H} \mathbf{w}_{i+1} \|_2}{\| \mathbf{w}_i \|_2 \|  \mathbf{w}_{i+1} \|_2 },
\end{equation}
where $\|\cdot\|_2$ denotes the L2 norm, and $\mathbf{w}_i$ and $\mathbf{w}_{i+1}$ are the dominant eigenvectors of the $i$-th and $i+1$-th subcarrier groups, respectively. Note that the closer the $\tilde{\rho}$ is to $1$, the more correlated the adjacent subcarrier groups are, while the closer the $\tilde{\rho}$ is to $0$, the less correlated the adjacent subcarrier groups are. Then, the average frequency domain correlations over all adjacent subcarrier groups and collected samples, denoted by $\overline{\tilde{\rho}}$, can be defined as
\begin{equation}
\overline{\tilde{\rho}} = \frac{1}{N_{col} (N_{grp} - 1)} \sum_{j=1}^{N_{col}} \sum_{i=1}^{N_{grp} - 1} \tilde{\rho}_j(i,i+1),
\end{equation}
where $N_{col}$ is the number of collected samples, $N_{grp}$ is the number of subcarrier groups.

Table \ref{fre cor} shows the average frequency domain correlations among adjacent subcarrier groups over all collected samples with different group granularities under the indoor and outdoor scenarios. According to Table III, to retain high frequency domain correlations (e.g., $\rho > 0.95$) for CSI feedback and knowledge-driven data augmentation, the group granularity of the proposed framework is set as $N_{gr} = 16$. Therefore, the size of the CSI matrix to be fed back is $32\times64$.
}

\subsubsection{Experimental Settings}

\begin{table}[htbp]
\centering
\caption{System setup for numerical experiments.}
\label{Sys. Set.}
\resizebox{\columnwidth}{!}{%
\begin{tabular}{c|c|c}
\hline
\textbf{System Parameter} & \textbf{Setting (Indoor)}        & \textbf{Setting (Outdoor)}        \\ \hline
\textbf{Scenario} & 
\makecell[c]{Indoor conference room scenario, \\$10$ m $\times$ $11$ m $\times$ $3$ m} & 
\makecell[c]{Outdoor scenario of two streets and \\one intersection} \\ \hline
\textbf{Active BS}                   & BS1                          & BS1                          \\ \hline
\textbf{BS antenna setup}            & ULA with $N_T=32$            & ULA with $N_T=32$            \\ \hline
\textbf{Height of BS}                & $2$ m                        & $6$ m                        \\ \hline
\textbf{UE antenna setup}            & ULA with $N_R=4$             & ULA with $N_R=4$             \\ \hline
\textbf{Operating frequency band}    & $60$ GHz                     & $3.5$ GHz                    \\ \hline
\textbf{System bandwidth}            & $100$ MHz                    & $100$ MHz                    \\ \hline
\textbf{Number of channel paths}     & $25$                         & $25$                         \\ \hline
\textbf{Number of subcarriers}  & $1024$                       & $1024$                       \\ \hline
\textbf{Group granularity}     & $16$                         & $16$                        \\ \hline
\textbf{Samples for training}     & \{$50$, $150$, $250$, $350$, $450$\}                         & \{$100$, $300$, $500$, $700$, $900$\} \\ \hline
\textbf{Samples for validation}     & $50$                         & $100$ \\ \hline
\textbf{Samples for testing}     & $4,500$                         & $9,000$                        \\ \hline
\end{tabular}
}
\end{table}

The number of feedback bits $B$ is set to $64$, $128$, $256$, $512$, or $1024$ for comprehensive performance comparison. For a fair comparison, all methods exploit $2$-bit uniform quantization. General hyper-parameter settings for the proposed IEFSF are shown in Table \ref{Hyper}. {Specifically, we set $\lambda_{1} = \lambda_{2} = 10$ following \cite{gulrajani2017improved} for training the EGAN.}

\begin{table}[htbp]
\centering
\caption{General hyper-parameter settings for the proposed IEFSF.}
\label{Hyper}
\resizebox{\columnwidth}{!}{%
\begin{tabular}{c|c|c}
\hline
\textbf{Component}             & \textbf{Hyper-parameter}                 & \textbf{Setting}                        \\ \hline
\multirow{6}{*}{\textbf{Compression   and Reconstruction Modules}}  & \textbf{MHSA layer dimension}     & $64$ \\ \cline{2-3} 
                               & \textbf{FF layer dimension}              & $256$                                   \\ \cline{2-3} 
                               & \textbf{attention head}                  & $4$                                     \\ \cline{2-3} 
                               & \textbf{dropout rate}                    & $0.1$                                   \\ \cline{2-3} 
                               & {$N_{com}$}                       & $2$                                     \\ \cline{2-3} 
                               & {$N_{rec}$}                       & $2$                                     \\ \hline
\multirow{2}{*}{\textbf{Quantization   and Dequantization Modules}} & \textbf{quantization bit   width} & $2$  \\ \cline{2-3} 
                               & \textbf{feedback bits}                   & \{$64$, $128$, $256$,   $512$, $1024$\} \\ \hline
\multirow{2}{*}{\textbf{Frequency   Extrapolation Network}}         & \textbf{linear layer   dimension} & $64$ \\ \cline{2-3} 
                               & \textbf{number of extrapolation modules} & $16$                                    \\ \hline
\multirow{9}{*}{\textbf{EGAN}} & {$l_z$}                           & $128$                                   \\ \cline{2-3} 
                               & {$d_{G}$}                         & $512$                                   \\ \cline{2-3} 
                               & {$d_{D1}$}                        & $32$                                    \\ \cline{2-3} 
                               & {$d_{D2}$}                        & $32$                                    \\ \cline{2-3} 
                               & {$N_G$}                           & $5$                                     \\ \cline{2-3} 
                               & {$N_{D1}$}                        & $5$                                     \\ \cline{2-3} 
                               & {$N_{D2}$}                        & $5$                                     \\ \cline{2-3} 
                               & $\lambda_1$                        & $10$                                     \\ \cline{2-3} 
                               & $\lambda_2$                        & $10$                                     \\ \hline
\end{tabular}
}
\end{table}

\begin{table}[htbp]
\centering
\caption{Quantified frequency domain correlations among adjacent subcarrier groups with different group granularities.}
\label{fre cor}
\begin{tabular}{c|c|c}
\hline
\textbf{Granularity} & \textbf{$\overline{\tilde{\rho}}$ (Indoor)} & \textbf{$\overline{\tilde{\rho}}$ (Outdoor)} \\ \hline
1                    & 1.0000                   & 0.9996                    \\ \hline
2                    & 1.0000                   & 0.9985                    \\ \hline
4                    & 1.0000                   & 0.9948                    \\ \hline
8                    & 0.9999                   & 0.9837                    \\ \hline
16                   & 0.9995                   & 0.9545                    \\ \hline
32                   & 0.9980                   & 0.9210                    \\ \hline
64                   & 0.9933                   & 0.9140                    \\ \hline
128                  & 0.9845                   & 0.8878                    \\ \hline
256                  & 0.9864                   & 0.8282                    \\ \hline
\end{tabular}
\end{table}

\subsubsection{Evaluation Metrics}

The average GCS $\bar{\rho}$ is adopted to evaluate the CSI feedback performance, which is defined as
\begin{equation}
\bar{\rho} = \frac{1}{N_{test}} \sum_{j=1}^{N_{test}} \bar{\rho}_j,
\end{equation}
where $N_{test}$ is the number of samples for testing, $\bar{\rho}_j$ is the subcarrier-level average GCS of the $j$-th testing sample, which is defined as
\begin{equation}
\bar{\rho}_j = \frac{1}{N_{c}} \sum_{i=1}^{N_{c}} \rho_i.
\end{equation}
Note that the closer the average GCS is to $1$, the better the CSI feedback accuracy is, while the closer the average GCS is to $0$, the worse the CSI feedback accuracy is.

In addition, the number of floating-point multiply-accumulate operations (`FLOPs') and the number of parameters (`Params') are adopted to evaluate the time complexity and space complexity of the proposed framework and the baselines, respectively.

\subsection{Performance Evaluation of Proposed Data Augmentation Methods for Few-Shot CSI Feedback}

We here evaluate the performance of the proposed knowledge-driven data augmentation method and AIGC-based data augmentation method for few-shot CSI feedback.

According to the proposed knowledge-driven data augmentation method, at most ${(C_{N_{sgrp}}^1)}^{N_{grp}}$ augmented samples can be generated with given subgroup granularity $N_{sgrp}$. However, these augmented samples are basically generated from limited collected samples relying on frequency domain correlations, which therefore lack of sample diversity. Considering the training cost and the model's generalization capability, in the experiments, we only generate samples $\widetilde{\mathbf{W}}_\mathbf{m} = [\widetilde{\mathbf{w}}_{1,m_1}, \widetilde{\mathbf{w}}_{2,m_2}, \ldots, \widetilde{\mathbf{w}}_{N_{grp},m_{N_{grp}}}]$ with $m_1=m_2= \ldots = m_{N_{grp}} = k$, and $k \in \{1, 2, \ldots, N_{sgr}\}$ from the knowledge-driven data augmentation method. Specifically, $N_{sgr}$ is set to $1$, $2$, $4$, and $8$ at the given group granularity $N_{gr} = 16$ to generate in total $30$ augmented samples for each collected sample.\footnote{The number of augmented samples is calculated by $16/1 + 16/2 + 16/4 + 16/8 = 30$.}

When applying the proposed AIGC-based data augmentation method, augmented data samples from the knowledge-driven data augmentation method are first generated and used to train the EGAN. After that, $20,000$ samples and $40,000$ samples are generated with trained EGANs for the indoor scenario and outdoor scenario, respectively.

\begin{figure}[htbp]
\centering
\begin{minipage}[t]{0.4\textwidth}
\centering
\includegraphics[width=\textwidth]{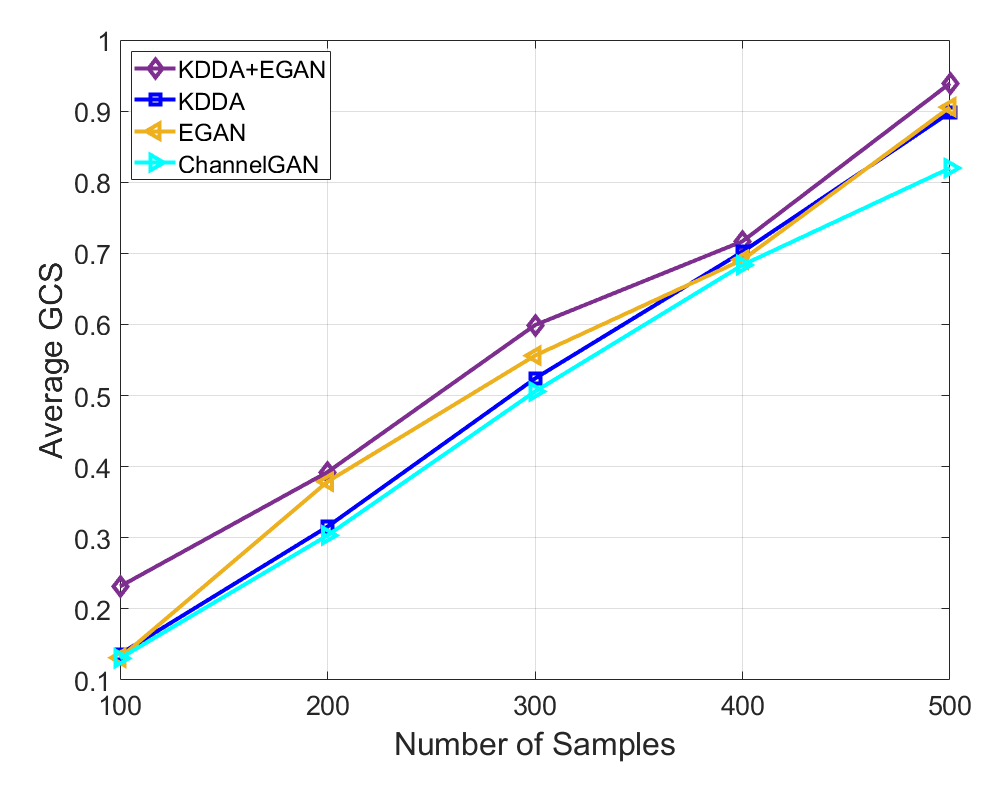}
\caption{Average GCS of the proposed data augmentation methods and the ChannelGAN-based method versus the numbers of collected samples under the indoor scenario.}
\label{I3_Aug}
\end{minipage}
\begin{minipage}[t]{0.4\textwidth}
\centering
\includegraphics[width=\textwidth]{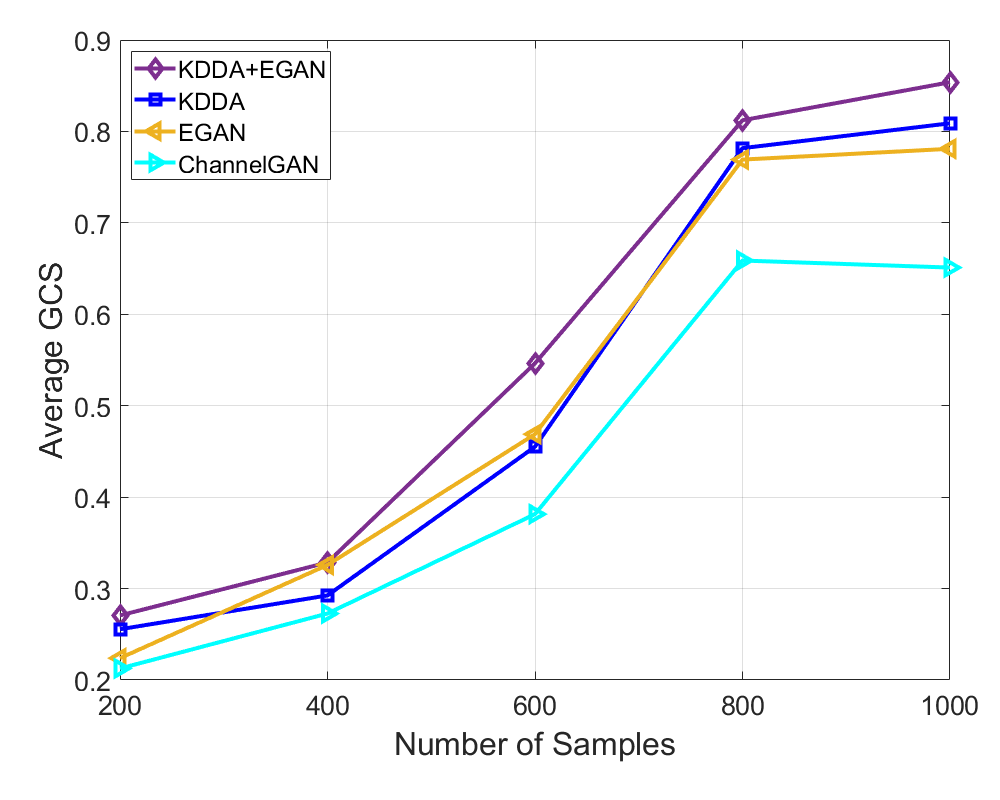}
\caption{Average GCS of the proposed data augmentation methods and the ChannelGAN-based method versus the numbers of collected samples under the outdoor scenario.}
\label{O1_Aug}
\end{minipage}
\end{figure}

The performance of the proposed knowledge-driven data augmentation method and the AIGC-based data augmentation method under indoor and outdoor scenarios is shown in Fig. \ref{I3_Aug} and Fig. \ref{O1_Aug}. In these figures, the horizontal axis represents the summation of training and validation samples (i.e., the number of collected samples), while the vertical axis represents the average GCS $\bar{\rho}$. For readability, these two data augmentation methods are denoted with `KDDA' and `EGAN' in the figures. And the feedback bits $B=256$. As depicted in Fig. \ref{I3_Aug} and Fig. \ref{O1_Aug}, the proposed methods show comparable performance in augmenting few-shot CSI feedback with only $100$ - $500$ collected samples for the indoor scenario and only $200$ - $1,000$ collected samples for the outdoor scenario. The knowledge-driven data augmentation method is applicable for scenarios with very few collected samples (e.g., only $100$-$200$ samples) while the AIGC-based method exhibits better performance when there are several hundred samples. Additionally, from Fig. \ref{I3_Aug} and Fig. \ref{O1_Aug}, by combining the two data augmentation methods (denoted by `KDDA+EGAN'), the proposed framework can achieve a high CSI feedback accuracy with only several hundred collected samples.


To further validate the effectiveness of the proposed AIGC-based data augmentation method, we compare it with the existing state-of-the-art (SOTA) AIGC-based data augmentation method for CSI feedback, i.e., the ChannelGAN-based method\cite{xiao2022channelgan}. For fare comparison, $20,000$ samples and $40,000$ samples are generated with the ChannelGAN for the indoor scenario and outdoor scenario, respectively. From Fig. \ref{I3_Aug}, Fig. \ref{O1_Aug}, and Table \ref{GANComplexity}, the EGAN-based method outperforms the ChannelGAN-based method given various numbers of collected samples under the scenarios, and the EGAN is much more efficient than the ChannelGAN. This is mainly due to the following three reasons: 1) In comparison to the ChannelGAN which learns to model the channel, the proposed EGAN directly learns to generate eigenvector-based CSI matrices, making it much more computationally efficient; 2) Thanks to the proposed two-scale generator and discriminator architecture, the EGAN learns both the distribution of full-dimensional CSI matrices and the distribution of low-dimensional CSI matrices, leading to more accurate distribution learning from limited collected samples; 3) With the assistance of the knowledge-driven data augmentation method, the EGAN can be trained with a high generalization capability.

Considering the performance and computational complexity of the proposed data augmentation methods,\footnote{According to Section \ref{Sec. DataAug}-A, the proposed knowledge-driven data augmentation method requires only several matrix multiplications to generate an augmented sample, and is thus much more computationally efficient and easier to implement than the proposed AIGC-based data augmentation method that requires the training of the EGAN and sample generation with the trained EGAN.} we conclude that the proposed knowledge-driven data augmentation method is more suitable for scenarios with very few collected samples (e.g., only $100$-$200$ samples). If there are several hundred collected samples and the training computational complexity is not a significant issue, the proposed AIGC-based data augmentation method can be trained well to provide better augmentation performance. Moreover, by combining the two data augmentation methods, a dataset with abundant and diverse samples can be formed to further improve CSI feedback accuracy under few-shot scenarios.

\begin{table}[htbp]
\centering
\caption{Computational complexity comparison of the proposed EGAN and the ChannelGAN.}
\label{GANComplexity}
\begin{tabular}{c|cc|cc}
\hline
\multirow{2}{*}{\textbf{Component}} & \multicolumn{2}{c|}{\textbf{ChannelGAN}}              & \multicolumn{2}{c}{\textbf{EGAN}}                    \\ \cline{2-5} 
                                    & \multicolumn{1}{c|}{\textbf{Params}} & \textbf{FLOPs} & \multicolumn{1}{c|}{\textbf{Params}} & \textbf{FLOPs} \\ \hline
\textbf{Generator}         & \multicolumn{1}{c|}{$19.51$ M} & $31.78$ G & \multicolumn{1}{c|}{$\bm{3.05}$ M} & $\bm{78.06}$ M \\ \hline
\textbf{Discriminator (1)} & \multicolumn{1}{c|}{$17.50$ M} & $2.15$ G  & \multicolumn{1}{c|}{$\bm{2.79}$ M} & $\bm{0.51}$ M  \\ \hline
\textbf{Discriminator 2}   & \multicolumn{1}{c|}{/}       & /       & \multicolumn{1}{c|}{$\bm{2.79}$ M} & $\bm{0.03}$ M  \\ \hline
\textbf{Whole Network}     & \multicolumn{1}{c|}{$37.01$ M} & $33.93$ G & \multicolumn{1}{c|}{$\bm{8.63}$ M} & $\bm{78.60}$ M \\ \hline
\end{tabular}
\end{table}

\subsection{Performance Comparison Among Proposed IEFSF and Baselines Under Few-Shot Scenarios}

{The proposed IEFSF is compared with the following learning-based baselines: 1) CsiNet+\cite{guo2020convolutional}, 2) bi-ImCsiNet\cite{chen2022deep} 3) TransNet\cite{cui2022transnet}, and 4) SampleDL\cite{wang2021compressive} under few-shot scenarios. The proposed IEFSF adopts the ``KDDA+EGAN'' data augmentation to tackle the challenge of lacking collected samples, while the baselines are trained with originally collected samples. In addition, we further compare the proposed IEFSF with conventional non-learning-based CSI feedback methods, including the random vector quantization (RVQ) based CSI feedback method \cite{jindal2006mimo} and the 5G NR Type I codebook-based CSI feedback method \cite{3gpp2020ts38214}. Note that since the RVQ-based method and the Type I codebook-based method do not learn from CSI samples, their performance does not change with the number of collected samples.} For fair comparisons, all baselines are adapted for eigenvector-based CSI feedback.

\begin{figure}[htbp] 
\centering
\begin{minipage}[t]{0.4\textwidth}
\centering
\includegraphics[width=\textwidth]{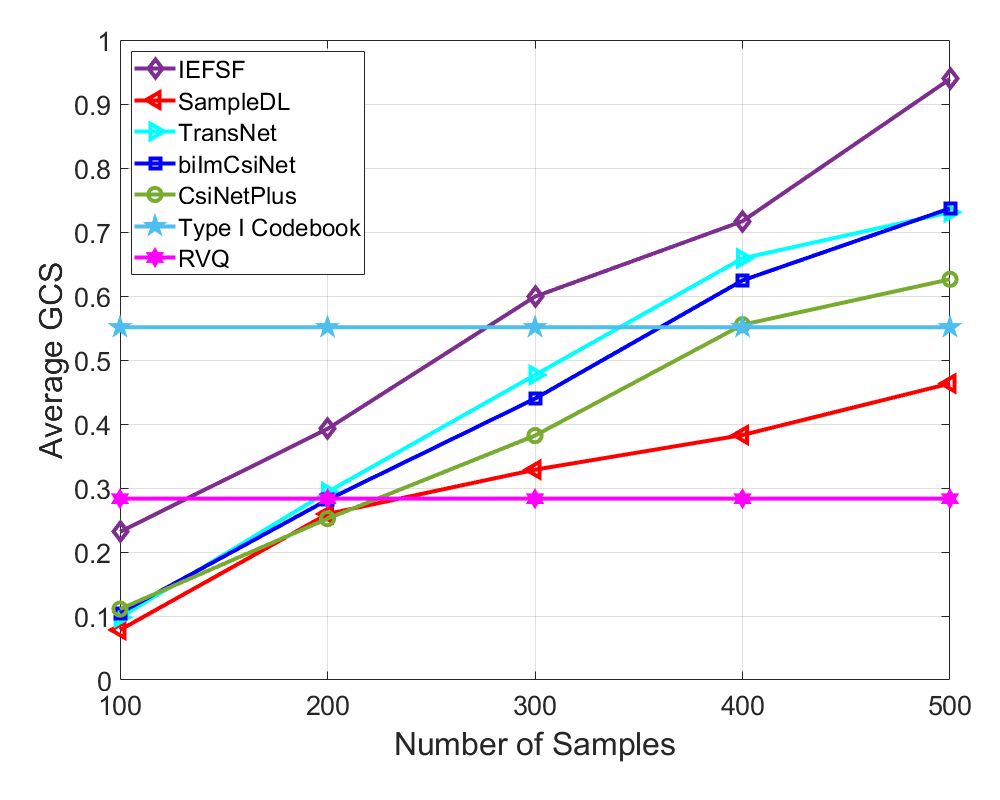}
\caption{Performance comparison of the proposed IEFSF and baselines versus the numbers of collected samples under the indoor scenario.}
\label{I3_GCS_N}
\end{minipage}
\begin{minipage}[t]{0.4\textwidth}
\centering
\includegraphics[width=\textwidth]{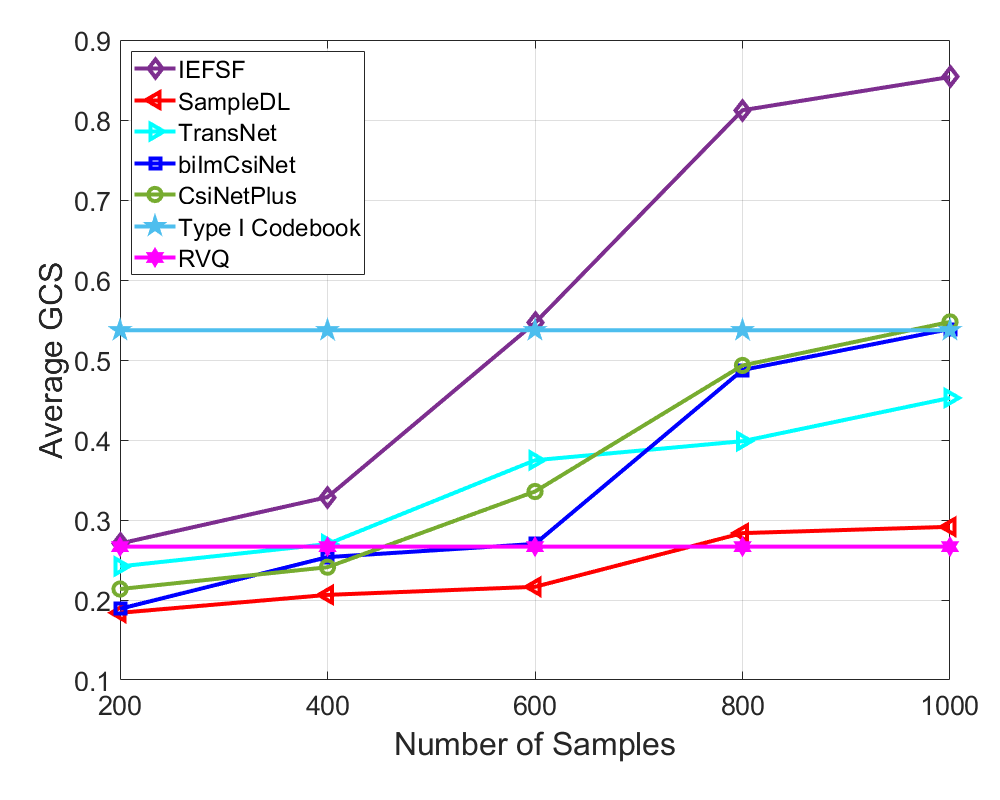}
\caption{Performance comparison of the proposed IEFSF and baselines versus the numbers of collected samples under the outdoor scenario.}
\label{O1_GCS_N}
\end{minipage}
\end{figure}

We first evaluate the impact of the number of collected samples with a fixed number of feedback bits $B=256$. As depicted in Fig. \ref{I3_GCS_N} and Fig. \ref{O1_GCS_N}, the IEFSF and the SampleDL show the best and the worst CSI feedback performance under both indoor and outdoor scenarios given few collected samples, respectively. Specifically, under the system configuration of $32$ antennas and $100$ MHz bandwidth with $1024$ subcarriers, the IEFSF can achieve up to $0.94$ average GCS with only $500$ collected samples for the indoor scenario and up to $0.85$ average GCS with only $1,000$ collected samples for the outdoor scenario. This demonstrates the superiority of the proposed IEFSF for few-shot CSI feedback. {Moreover, it can be seen from Fig. \ref{I3_GCS_N}  that the proposed IEFSF outperforms the RVQ-based CSI feedback method with only $200$ collected samples and outperforms the Type I codebook-based CSI feedback method with only $300$ collected samples under the indoor scenario. From Fig. \ref{O1_GCS_N}, the proposed IEFSF outperforms the RVQ-based CSI feedback method with only $200$ collected samples and outperforms the Type I codebook-based CSI feedback method with only $600$ collected samples under the outdoor scenario. These results indicate that only a few collected samples are required by the proposed IEFSF to surpass non-learning baselines and further demonstrate the practicality of our proposed IEFSF for adapting to new scenarios.}

\begin{figure}[htbp] 
\centering
\begin{minipage}[t]{0.4\textwidth}
\centering
\includegraphics[width=\textwidth]{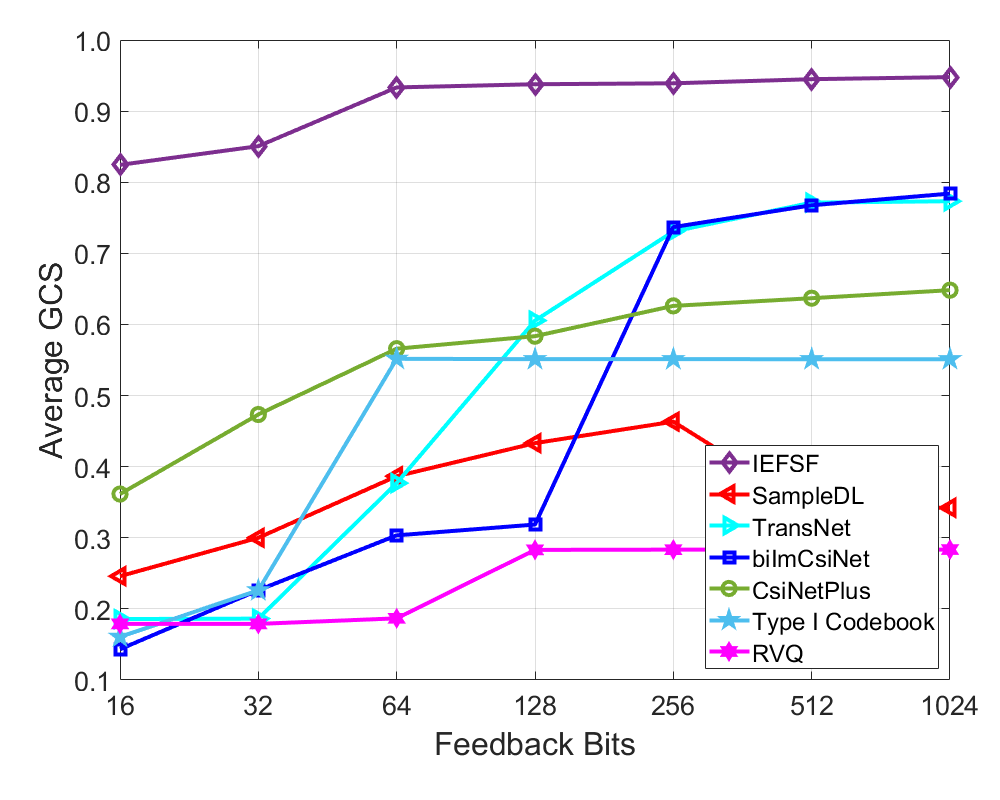}
\caption{Average GCS of the proposed IEFSF and baselines versus the numbers of feedback bits under the indoor scenario.}
\label{I3_GCS_FB}
\end{minipage}
\begin{minipage}[t]{0.4\textwidth}
\centering
\includegraphics[width=\textwidth]{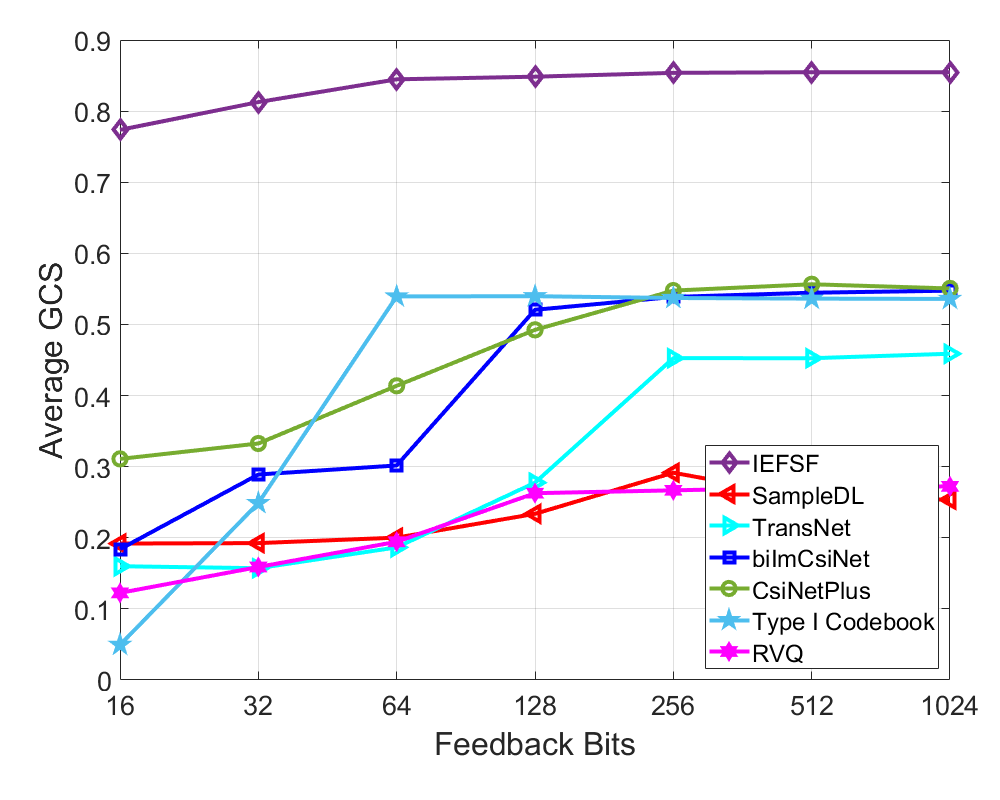}
\caption{Average GCS of the proposed IEFSF and baselines versus the numbers of feedback bits under the outdoor scenario.}
\label{O1_GCS_FB}
\end{minipage}
\end{figure}

The impact of feedback bits on the CSI feedback performance is also investigated. The number of collected samples is fixed at $500$ and $1000$ for the indoor scenario and the outdoor scenario, respectively in these experiments. It can be observed from Fig. \ref{I3_GCS_FB} and Fig. \ref{O1_GCS_FB} that more feedback bits bring about higher feedback accuracy. However, as the number of feedback bits increases to a relatively large value (e.g., $256$), the performance improvement becomes limited, showing a marginal effect. Besides, increasing the number of feedback bits will also lead to increasing computational complexity. Therefore, a proper setting of feedback bits deserves a dedicated selection in practical applications. {In addition, as shown in Fig. \ref{I3_GCS_FB} and Fig. \ref{O1_GCS_FB}, the IEFSF respectively achieves around $0.88$ and $0.77$ average GCS under the indoor and outdoor scenarios with only $16$ feedback bits. As a comparison, the learning-based baselines merely achieve at most $0.78$ and $0.55$ average GCS and the conventional non-learning-based baselines merely achieve at most $0.55$ and $0.54$ average GCS under both scenarios with $1024$ feedback bits. This indicates the proposed IEFSF can achieve high CSI feedback accuracy with much lower feedback overhead (i.e., up to $64$ times lower feedback overhead than the baselines).} It is worth mentioning that performance degradation of SampleDL can be observed when the number of feedback bits increases from $256$ onwards. This is because the SampleDL adopts a two-step training scheme, i.e., training the CSI feedback network first and then training the interpolating network. Given a large number of feedback bits while training the CSI feedback network with few training samples, the trained CSI feedback network is prone to overfitting to the first-stage CSI feedback. As a consequence, the interpolating network can not be trained properly from biased outputs from the trained CSI feedback network. Different from the SampleDL, the proposed IEFSF adopts joint training of CSI feedback and interpolation, leading to a stable feedback accuracy improvement with the increase of feedback bits.

\begin{figure}[htbp] 
\centering
\begin{minipage}[t]{0.4\textwidth}
\centering
\includegraphics[width=\textwidth]{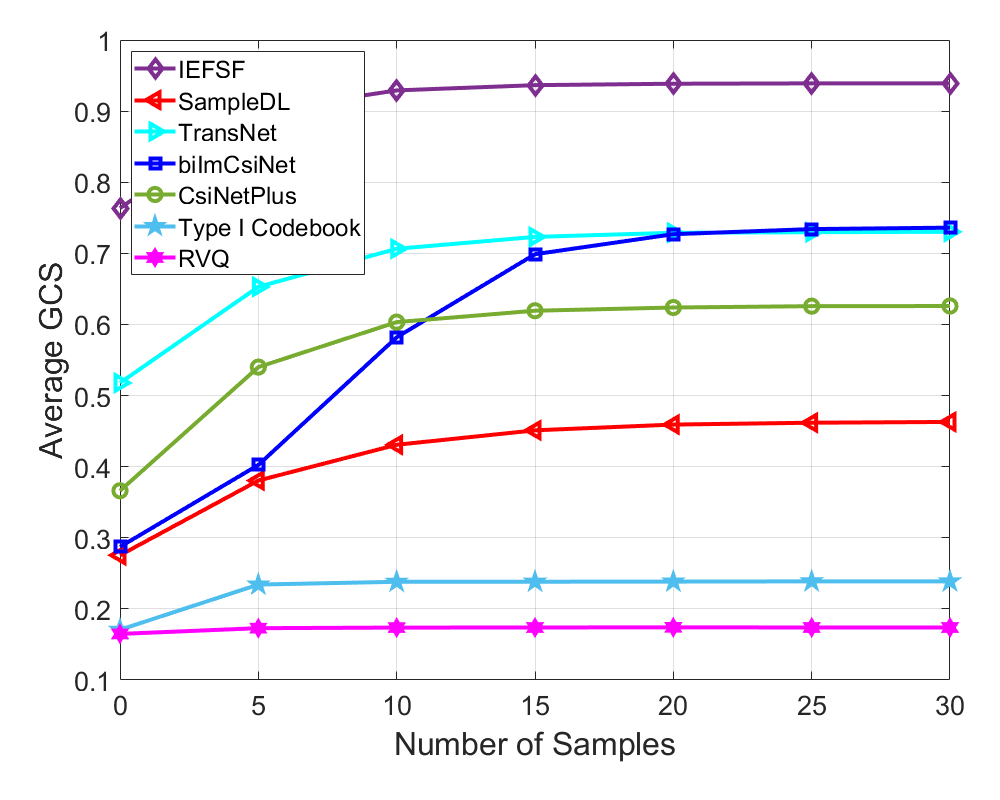}
\caption{Average GCS of the proposed IEFSF and baselines versus SNRs under the indoor scenario.}
\label{I3_SNR}
\end{minipage}
\begin{minipage}[t]{0.4\textwidth}
\centering
\includegraphics[width=\textwidth]{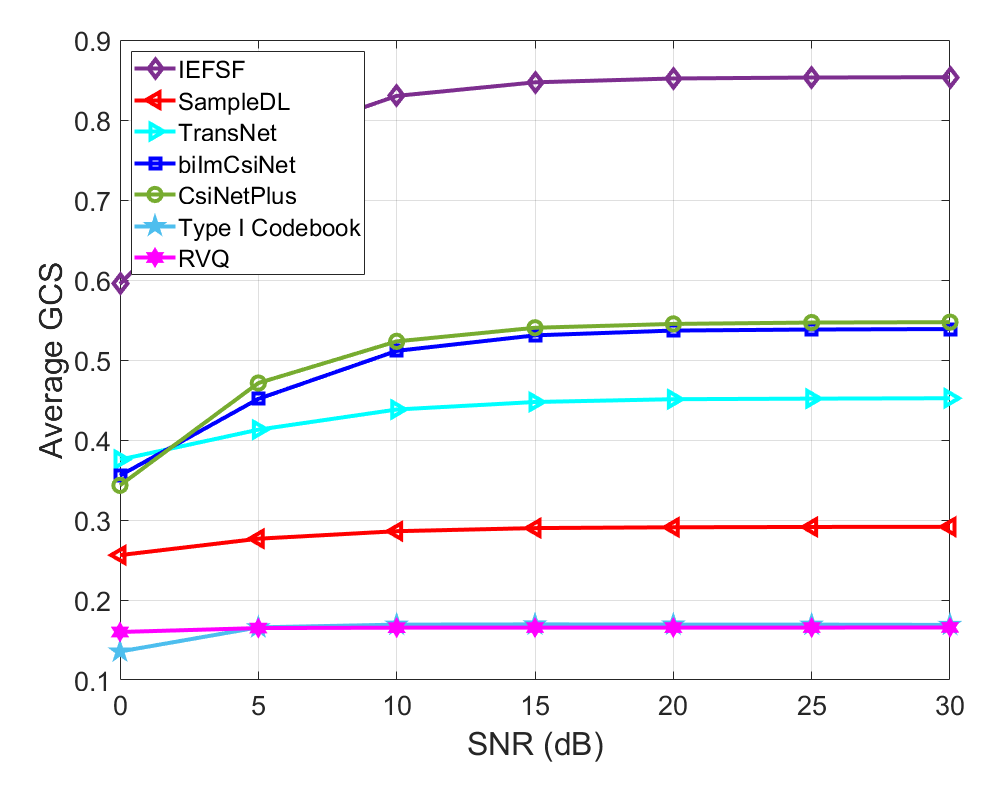}
\caption{Average GCS of the proposed IEFSF and baselines versus SNRs under the outdoor scenario.}
\label{O1_SNR}
\end{minipage}
\end{figure}

{We further investigate the performance of the proposed IEFSF and baselines versus various SNRs. The number of feedback bits is fixed to $B=256$ and the number of collected samples is fixed at $500$ and $1000$ for the indoor scenario and the outdoor scenario, respectively. As shown in Fig. \ref{I3_SNR} and Fig. \ref{O1_SNR}, the proposed IEFSF outperforms the baselines either with a low SNR or with a high SNR under both scenarios and higher SNRs facilitate higher feedback accuracy in terms of the average GCS. In the SNR regime higher than $20$ dB, the proposed IEFSF and all learning-based baselines achieve a performance similar to that in noise-free cases. It is worth noting that the non-learning-based baselines, i.e., the RVQ-based CSI feedback method and the Type I codebook-based CSI feedback method, are very sensitive to the noise since they directly feed back the indices of codewords to the BS, therefore they can not achieve a performance similar to that in noise-free cases even with a $30$ dB SNR. This further demonstrates the robustness of learning-based CSI feedback methods over conventional non-learning-based methods.}

Finally, we compare the computational complexity of the proposed IEFSF with learning-based baselines. The computational complexity of all schemes with a fixed setting of feedback bits $B=256$ is shown in Table \ref{Complexity}. From Table \ref{Complexity}, the IEFSF and the SampleDL show computational efficiency over common DL-based CSI feedback schemes, i.e., CsiNet+, bi-ImCsiNet, and TransNet. This is because the SampleDL and the proposed IEFSF greatly reduce the CSI dimension for feedback before the CSI compression, such that the computational complexity can be significantly reduced. Further comparing the proposed IEFSF with the SampleDL, the proposed IEFSF has a significantly smaller time complexity than the SampleDL (i.e., $25.1$ times smaller) at the cost of a slightly larger space complexity (i.e., $1.3$ times larger than the SampleDL).

\begin{table*}[htbp]
\centering
\caption{Computational complexity of the proposed IEFSF and baselines.}
\label{Complexity}
\resizebox{\textwidth}{!}{%
\begin{tabular}{c|cc|cc|cc|cc|cc}
\hline
\multirow{2}{*}{\textbf{Component}} &
  \multicolumn{2}{c|}{\textbf{CsiNet+}} &
  \multicolumn{2}{c|}{\textbf{bi-ImCsiNet}} &
  \multicolumn{2}{c|}{\textbf{TransNet}} &
  \multicolumn{2}{c|}{\textbf{SampleDL}} &
  \multicolumn{2}{c}{\textbf{IEFSF}} \\ \cline{2-11} 
 &
  \multicolumn{1}{c|}{\textbf{Params}} &
  \textbf{FLOPs} &
  \multicolumn{1}{c|}{\textbf{Params}} &
  \textbf{FLOPs} &
  \multicolumn{1}{c|}{\textbf{Params}} &
  \textbf{FLOPs} &
  \multicolumn{1}{c|}{\textbf{Params}} &
  \textbf{FLOPs} &
  \multicolumn{1}{c|}{\textbf{Params}} &
  \textbf{FLOPs} \\ \hline
\textbf{Encoder (UE)} &
  \multicolumn{1}{c|}{$8.39$ M} &
  $21.63$ M &
  \multicolumn{1}{c|}{$9.17$ M} &
  $818.28$ M &
  \multicolumn{1}{c|}{$8.66$ M} &
  $572.53$ M &
  \multicolumn{1}{c|}{$\bm{0.54}$ M} &
  $24.95$ M &
  \multicolumn{1}{c|}{$0.63$ M} &
  $\bm{7.37}$ M\\ \hline
\textbf{Decoder (BS)} &
  \multicolumn{1}{c|}{$8.48$ M} &
  $727.38$ M &
  \multicolumn{1}{c|}{$285.96$ M} &
  $285.97$ M &
  \multicolumn{1}{c|}{$8.73$ M} &
  $572.59$ M &
  \multicolumn{1}{c|}{$\bm{0.53}$ M} &
  $10.50$ M &
  \multicolumn{1}{c|}{$0.64$ M} &
  $\bm{7.37}$ M \\ \hline
\textbf{Extrapolation Network (BS)} &
  \multicolumn{1}{c|}{/} &
  / &
  \multicolumn{1}{c|}{/} &
  / &
  \multicolumn{1}{c|}{/} &
  / &
  \multicolumn{1}{c|}{$\bm{0.02}$ M} &
  $547.55$ M &
  \multicolumn{1}{c|}{$0.13$ M} &
  $\bm{8.46}$ M \\ \hline
\textbf{Whole Network} &
  \multicolumn{1}{c|}{$16.87$ M} &
  $749.01$ M &
  \multicolumn{1}{c|}{$295.13$ M} &
  $1104.25$ M &
  \multicolumn{1}{c|}{$17.39$ M} &
  $1145.12$ M &
  \multicolumn{1}{c|}{$\bm{1.09}$ M} &
  $583.00$ M &
  \multicolumn{1}{c|}{$1.40$ M} &
  $\bm{23.20}$ M \\ \hline
\end{tabular}
}
\end{table*}

\section{Conclusion} \label{Sec. Con.}
This paper proposed a low-overhead incorporation-extrapolation based few-shot CSI feedback framework for FDD massive MIMO systems, namely the IEFSF. Through the incorporation-extrapolation CSI feedback scheme, the UE fed the low-dimensional eigenvector-based CSI formed with the incorporation process back to the BS and then the BS recovered the full-dimensional eigenvector-based CSI via the extrapolation process, which can significantly reduce the feedback overhead and computational complexity. To alleviate the necessity of the extensive collected samples and enable few-shot CSI feedback, a knowledge-driven data augmentation method was first proposed for CSI data augmentation by exploiting frequency domain correlations, which was demonstrated to be effective and easy to implement especially. Then, an AIGC-based data augmentation method exploiting the novel EGAN architecture was further proposed to improve the diversity of augmented data samples, which was shown to be effective and much more computationally efficient than existing AIGC-based data augmentation methods. Extensive experimental results and computational complexity analysis demonstrated the effectiveness of the proposed IEFSF in reducing CSI feedback overhead and computational complexity. Remarkably, even with only several hundreds of collected samples, the IEFSF can achieve high CSI feedback accuracy under both indoor and outdoor scenarios.

\section{Acknowledgement}
This work was performed in part at SICC which is supported by SKL-IOTSC, University of Macau.

\bibliographystyle{IEEEtran}
\bibliography{refer}

\begin{IEEEbiography}
    [{\includegraphics[width=1in,height=1.25in,clip,keepaspectratio]{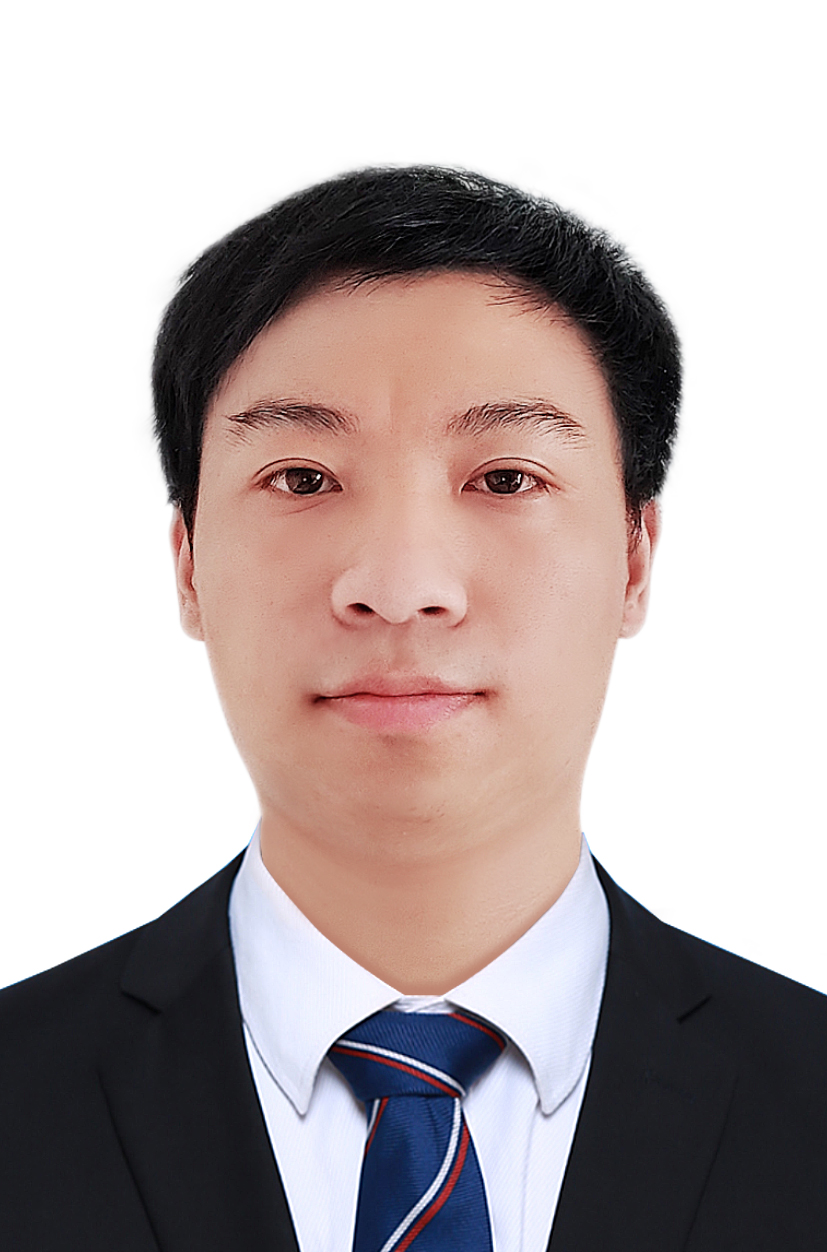}}]{Binggui Zhou} received the B.Eng. degree from Jinan University, Zhuhai, China, in 2018, and the M.Sc. degree from the University of Macau, Macao, China, in 2021, respectively. He is currently working toward the Ph.D. degree in Electrical and Computer Engineering with the University of Macau, Macao, China. He also serves as a Research Assistant with the School of Intelligent Systems Science and Engineering, Jinan University, Zhuhai, China. His research interests include artificial intelligence (AI), AI empowered wireless communications, and data mining.
\end{IEEEbiography}

\begin{IEEEbiography}
    [{\includegraphics[width=1in,height=1.25in,clip,keepaspectratio]{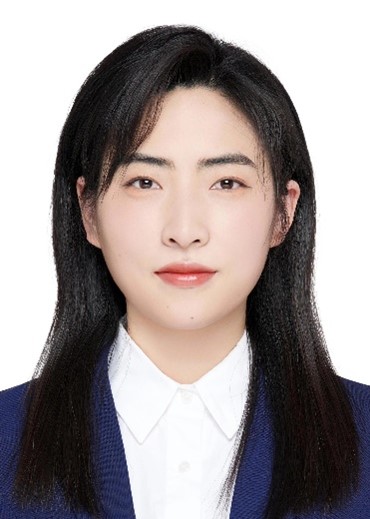}}]{Xi Yang} received the B.S., M.Eng. and Ph.D. degrees from Southeast University, Nanjing, China, in 2013, 2016 and 2019, respectively. From July 2020 to July 2022, she was a postdoctoral fellow with the State Key Laboratory of Internet of Things for Smart City, University of Macau, China. She is currently a Zijiang Young Scholar with the School of Communication and Electronic Engineering, East China Normal University, Shanghai, China. Her current research interests include extremely large aperture array (ELAA) systems, millimeter wave communications, and wireless communication system prototyping.
\end{IEEEbiography}

\begin{IEEEbiography}
    [{\includegraphics[width=1in,height=1.25in,clip,keepaspectratio]{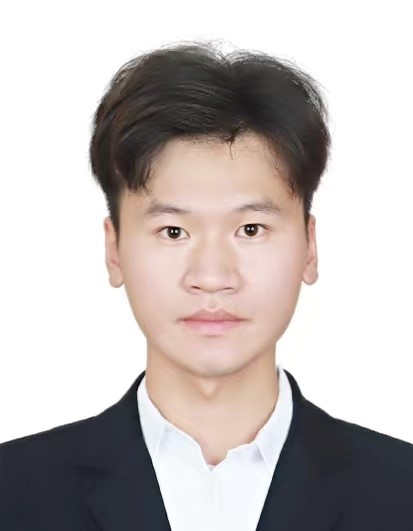}}]{Jintao Wang} received the B.Eng. degree in 2020 in communication engineering from Jilin University, Changchun, China. He is currently pursuing the Ph.D. degree with the State Key Laboratory of Internet of Things for Smart City and the Department of Electrical and Computer Engineering, University of Macau, Macau, China. His main research interests include RIS-aided communication, mmWave communication, transceiver design, and convex optimization.
\end{IEEEbiography}

\begin{IEEEbiography}
    [{\includegraphics[width=1in,height=1.25in,clip,keepaspectratio]{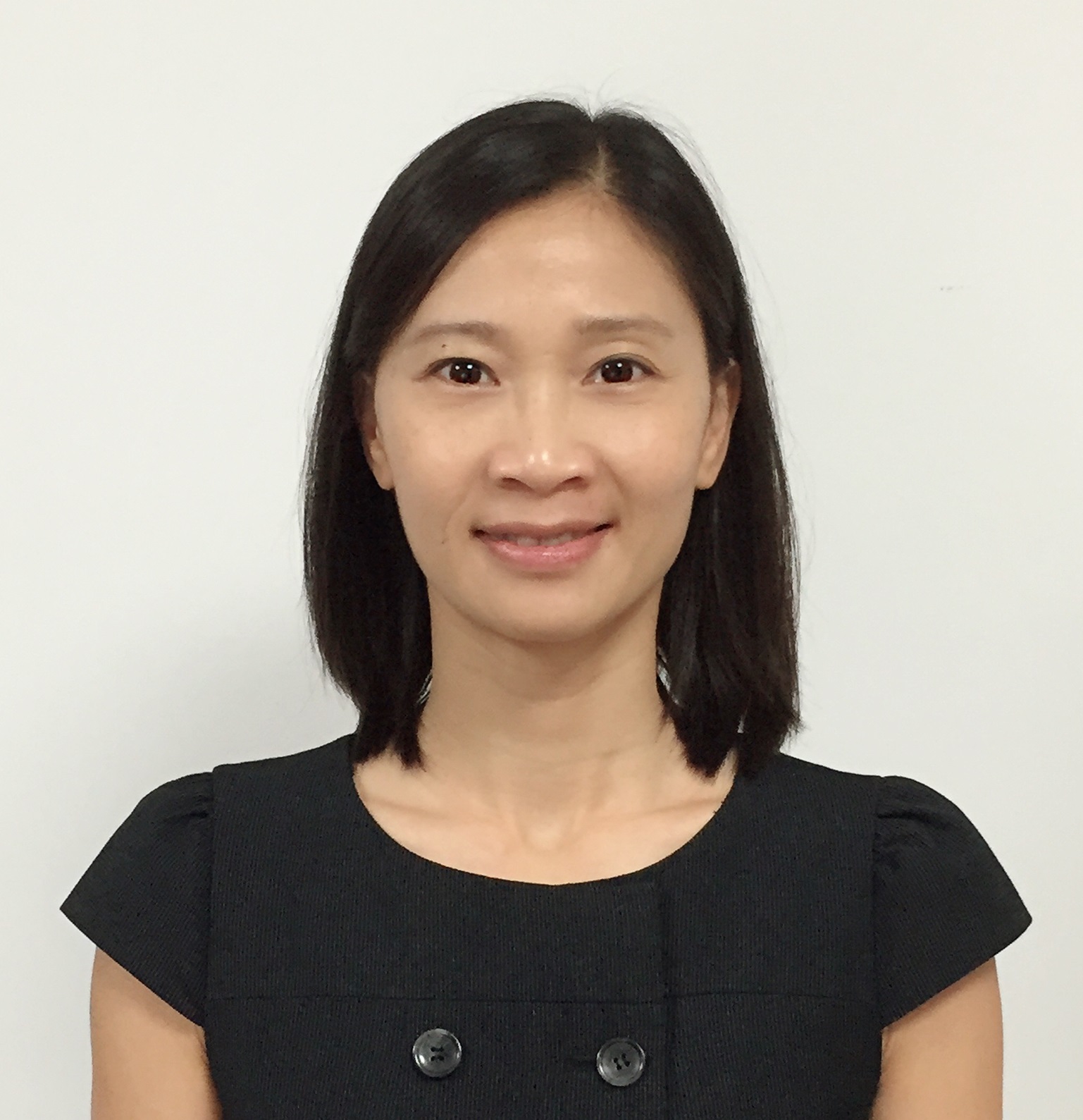}}]{Shaodan Ma} (Senior Member, IEEE) received the double Bachelor’s degrees in science and economics and the M.Eng. degree in electronic engineering from Nankai University, Tianjin, China, in 1999 and 2002, respectively, and the Ph.D. degree in electrical and electronic engineering from The University of Hong Kong, Hong Kong, in 2006. From 2006 to 2011, she was a post-doctoral fellow at The University of Hong Kong. Since August 2011, she has been with the University of Macau, where she is currently a Professor. Her research interests include array signal processing, transceiver design, localization, integrated sensing and communication, mmWave/THz communications, massive MIMO, and machine learning for communications. She was a symposium co-chair for various conferences including IEEE VTC2024-Spring, IEEE ICC 2021, 2019 \& 2016, IEEE GLOBECOM 2016, IEEE/CIC ICCC 2019, etc. She is an IEEE ComSoc Distinguished Lecturer in 2024-2025 and has served as an Editor for IEEE Wireless Communications (2024-present), IEEE Communications Letters (2023), Journal of Communications and Information Networks (2021-present), IEEE Transactions on Wireless Communications (2018-2023), IEEE Transactions on Communications (2018-2023), and IEEE Wireless Communications Letters (2017-2022).
\end{IEEEbiography}

\begin{IEEEbiography}
    [{\includegraphics[width=1in,height=1.25in,clip,keepaspectratio]{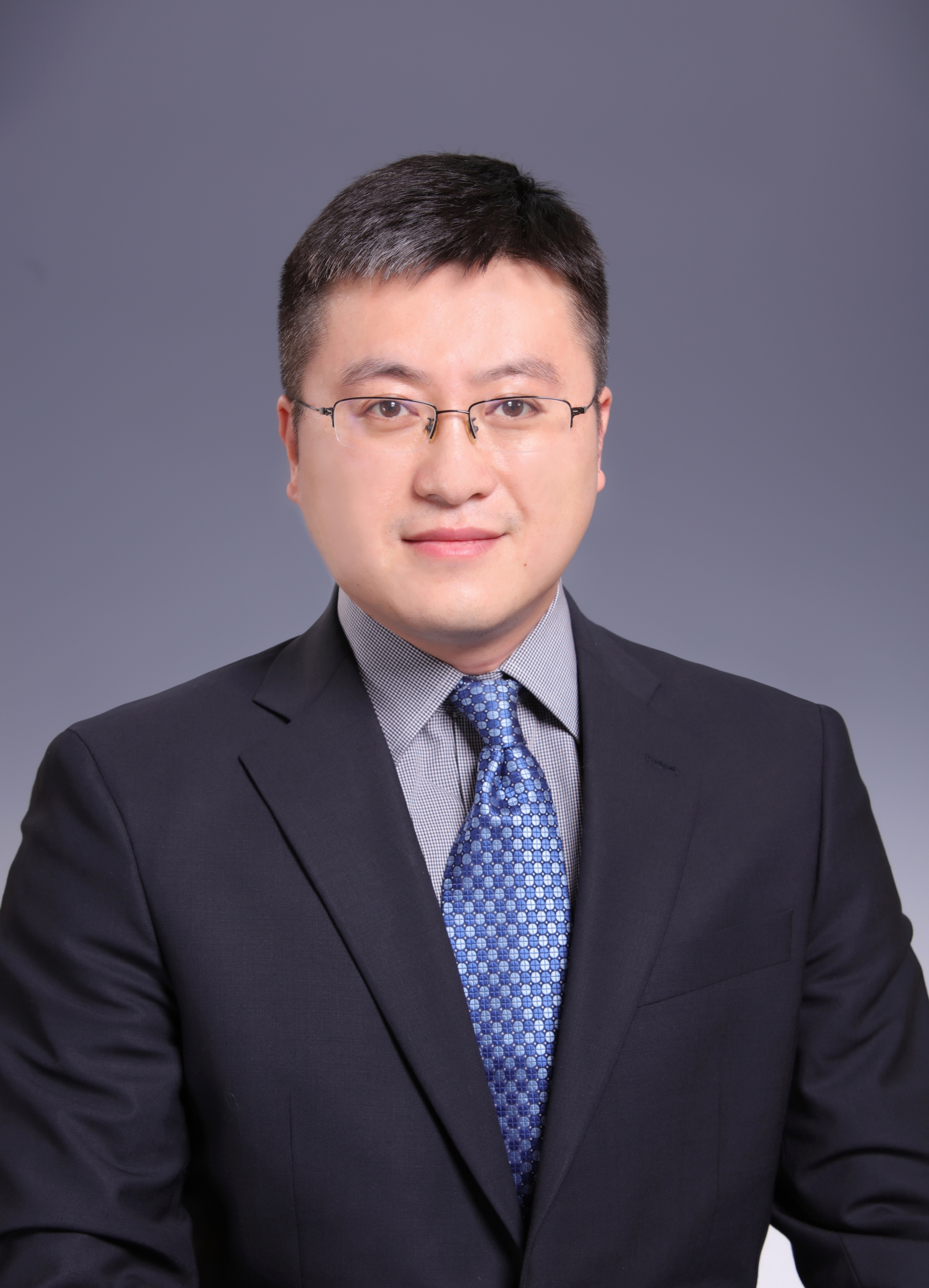}}] {Feifei Gao} (Fellow, IEEE) received the B.Eng. degree from Xi'an Jiaotong University, Xi'an, China in 2002, the M.Sc. degree from McMaster University, Hamilton, ON, Canada in 2004, and the Ph.D. degree from National University of Singapore, Singapore in 2007. Since 2011, he joined the Department of Automation, Tsinghua University, Beijing, China, where he is currently a tenured full professor. 

    Prof. Gao's research interests include signal processing for communications, array signal processing, convex optimizations, and artificial intelligence assisted communications. He has authored/coauthored more than 200 refereed IEEE journal papers and more than 150 IEEE conference proceeding papers that are cited more than 19000 times in Google Scholar. Prof. Gao has served as an Editor of IEEE Transactions on Wireless Communications, IEEE Journal of Selected Topics in Signal Processing (Lead Guest Editor), IEEE Transactions on Cognitive Communications and Networking, IEEE Signal Processing Letters (Senior Editor), IEEE Communications Letters (Senior Editor), IEEE Wireless Communications Letters, and China Communications. He has also served as the symposium co-chair for 2019 IEEE Conference on Communications (ICC), 2018 IEEE Vehicular Technology Conference Spring (VTC), 2015 IEEE Conference on Communications (ICC), 2014 IEEE Global Communications Conference (GLOBECOM), 2014 IEEE Vehicular Technology Conference Fall (VTC), as well as Technical Committee Members for more than 50 IEEE conferences.
\end{IEEEbiography}

\begin{IEEEbiography}
    [{\includegraphics[width=1in,height=1.25in,clip,keepaspectratio]{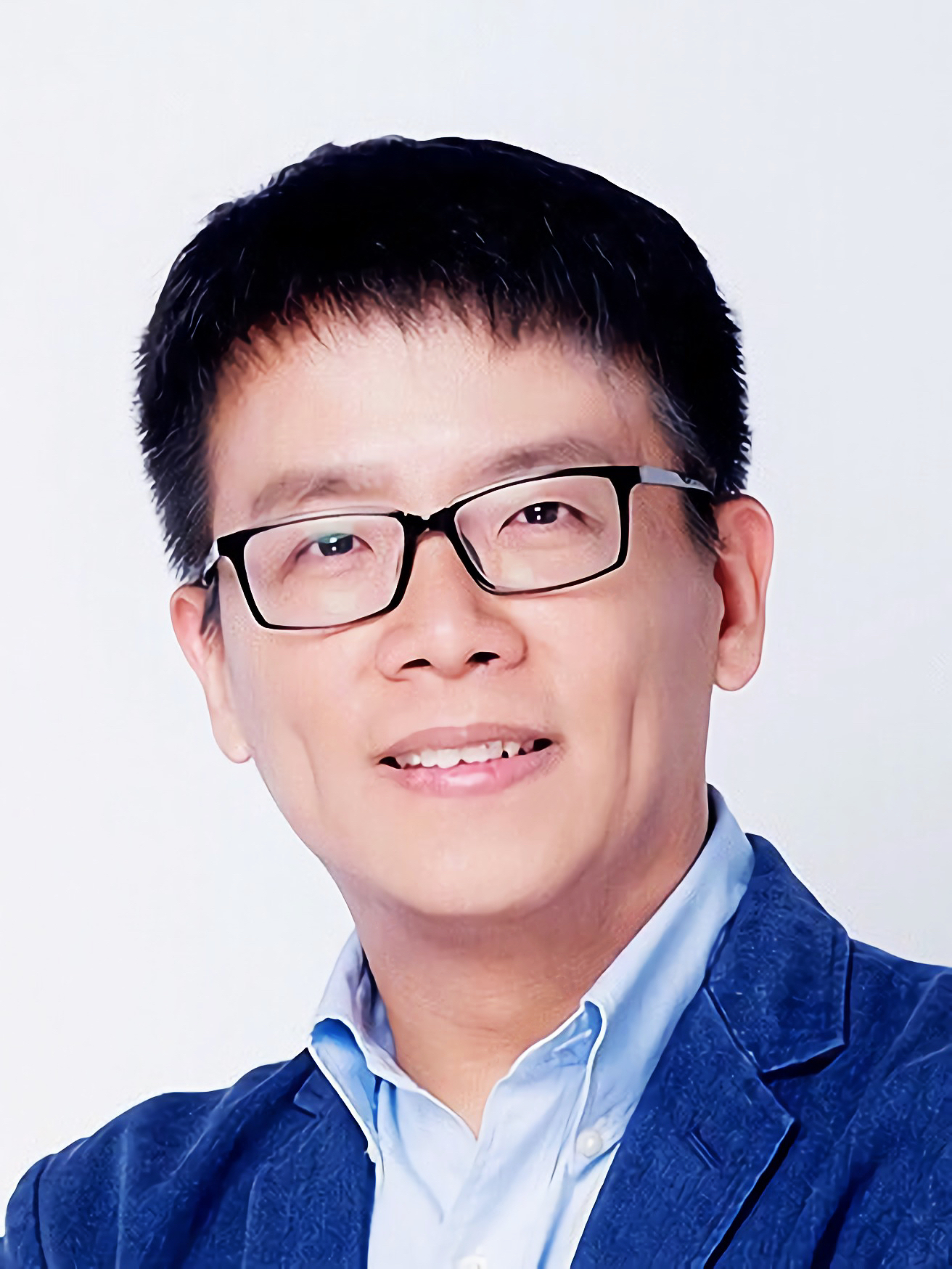}}]{Guanghua Yang} (Senior Member, IEEE) received his Ph.D. degree in electrical and electronic engineering from the University of Hong Kong in 2006. From 2006 to 2013, he served as post-doctoral fellow, research associate at the University of Hong Kong. Since April 2017, he has been with Jinan University, where he is currently a Full Professor in the School of Intelligent Systems Science and Engineering. His research interests are in the general areas of communications and networking.
\end{IEEEbiography}
\end{document}